\documentclass[aps,pra,twocolumn,superscriptaddress,floatfix]{revtex4-2}

\usepackage{amsmath}
\usepackage{amssymb}
\usepackage{graphicx}
\usepackage{bm}
\usepackage{booktabs}
\usepackage{siunitx}
\usepackage{multirow}
\usepackage[colorlinks=true,linkcolor=blue,citecolor=blue,urlcolor=blue]{hyperref}

\graphicspath{{figure/}}

\begin{document}

\title{Exchange and core-polarization effects on Rydberg-transition electric-dipole matrix elements in Rb and Cs}
\author{Shi-Cheng Yu}
\affiliation{Center for Theoretical Physics, School of Physics and Optoelectronic Engineering, Hainan University, Haikou 570228, China}
\author{Zhen-Xiang Zhong}
\email{zxzhong@hainanu.edu.cn}
\affiliation{Center for Theoretical Physics, School of Physics and Optoelectronic Engineering, Hainan University, Haikou 570228, China}
\author{Cheng-Bin Li}
\email{cbli@apm.ac.cn}
\affiliation{Innovation Academy for Precision Measurement Science and Technology, Chinese Academy of Sciences, Wuhan 430071, China}


\begin{abstract}
Microwave electric-field measurements with Rydberg atoms require accurate
Rydberg-transition electric-dipole matrix elements. We calculate these matrix
elements for Rb and Cs using a Dirac-Fock plus core-polarization
method, which core-valence exchange is treated explicitly and
core-valence correlation is represented by a cutoff core-polarization
potential. A smooth cutoff-radius model fitted to quantum-defect
energies reproduces the target energies with MHz-level residuals for states
up to $n=90$. Comparisons with model-potential matrix elements and internal
length-velocity consistency show that the $s$-$p$ transitions are generally
the most robust, whereas the $p$-$d$ and $d$-$f$ transitions show stronger
atom- and branch-dependent sensitivity to short-range modeling. The largest
relative differences arise from radial-integral cancellation, where
short-range phase changes are amplified by the small final matrix element.
These results provide a practical reliability assessment for
Rydberg-transition electric-dipole matrix elements used in field-sensing
applications.
\end{abstract}

\maketitle

\section{Introduction}

Rydberg atoms provide large electric-dipole ($E1$) moments and an atom-based 
route to SI-traceable microwave electric-field measurements \cite{RevModPhys.82.2313, 6910267, PhysRevA.101.053432}. In an
Autler-Townes measurement, the microwave electric field couples two Rydberg
levels and produces a spectroscopic splitting that is directly proportional to
the Rydberg-transition $E1$ matrix element \cite{Gordon, Holloway2014, Holloway2017}. With polarization
factors included in the projected matrix element, the relation may be written
as
\begin{equation}
  h\Delta\nu_{\mathrm{AT}} = |\mathbf{E}\cdot\mathbf{D}|,
  \label{eq:at_splitting}
\end{equation}
where $\Delta\nu_{\mathrm{AT}}$ is the measured splitting, $\mathbf{E}$ is the
microwave electric-field amplitude, and $\mathbf{D}$ is the projected 
Rydberg-transition $E1$ matrix element.  The $E1$
matrix element therefore enters the electric-field calibration directly.
As experimental spectra and field-readout methods improve, the reliability of
Rydberg-transition $E1$ matrix elements becomes a practical metrology question
rather than a detail of atomic-structure calculations \cite{Holloway2017, PhysRevApplied.21.044025, Simons2016}.

The main theoretical challenge is to describe a Rydberg wavefunction over a
wide range of length scales, from the near-core region to radii of order
$n^2 a_0$. Many-body perturbation theory (MBPT) and all-order methods can achieve 
high accuracy for low-lying states \cite{PhysRevA.69.022509, PhysRevA.94.012505}, but
applying them directly to Rydberg states is computationally
demanding. The radial box must be large, the wavefunctions become highly 
oscillatory, the
level spacing decreases rapidly, and finite-basis treatments require an
increasing number of basis functions, which rapidly increases the computational
cost \cite{PhysRevA.85.062709, Hansen_1993, Sapirstein_1996}. Practical calculations
of Rydberg spectra and matrix elements therefore often use a semiempirical
single-active-electron model potential \cite{PhysRevA.49.982}. Fitted to
experimental level energies, this potential provides an efficient baseline 
for Rydberg-atom calculations \cite{Yu_2025, PhysRevA.109.022810, PhysRevA.110.043114, PhysRevA.111.053120}.

An energy fit, however, does not directly constrain the radial matrix element.
The $E1$ matrix elements depend on the relative radial phases, node positions,
and amplitudes of the two orbitals, together with the cancellation pattern of
the radial integral \cite{MIGDALEK2020101355}. Short-range
differences in the effective potential can change the phase accumulated by the
Rydberg electron, thereby modifying the wavefunction in the outer classically
allowed region where the dominant contribution to the radial overlap integral
arises. This sensitivity is especially important when the final matrix element
results from a cancellation between larger positive and negative contributions.
In such cases, a modest change in the wavefunctions can produce a large
relative change in the $E1$ matrix element, even when the corresponding energy
shift is small.

The Dirac-Fock plus core-polarization method (DFCP) provides a complementary
framework for generating Rydberg orbitals. In this approach, the closed-shell
ionic core is described by the Dirac-Fock (DF) method, so that exchange with the
core is included explicitly at the DF level. Core-valence correlation
is represented by a cutoff core-polarization potential. This retains more
near-core many-electron structure than a purely single-active-electron model
potential, while remaining much less expensive than MBPT or all-order
calculations for Rydberg states.

Previous DFCP treatments have often used a B-spline basis, which provides a
finite and nearly complete basis for correlation calculations
\cite{PhysRevA.94.032503, PhysRevA.109.063115, cwhz-by5w, PhysRevA.87.042517}.
For Rydberg states, however, the spectrum obtained from such a finite and
nearly complete basis expansion contains many discretized-continuum and
pseudostate levels in the same energy region as the physical Rydberg states. 
In dense Rydberg manifolds, this can make the physical Rydberg
state difficult to identify and can limit the range over which individual
Rydberg states can be followed reliably. Existing DFCP calculations
of alkali Rydberg states have been demonstrated up to about \(n=50\)
\cite{PhysRevA.94.032503}.

In this work, we implement the DFCP method in a Adams-Moulton radial 
integration scheme for Rb and Cs Rydberg states. The target orbital is 
selected directly by its radial node number, avoiding the
need to identify it among the many pseudostate levels generated by a finite and
nearly complete basis expansion. A Newton procedure is used to determine
cutoff radii that reproduce quantum-defect energies, and a smooth fitted
cutoff-radius model is then used to calculate Rydberg states up to \(n=90\).
We use the resulting orbitals to assess the sensitivity of
Rydberg-transition \(E1\) matrix elements to exchange, core-polarization,
radial-integral cancellation, and gauge form. Although the method still relies
on fitted cutoff parameters, the explicit treatment of core exchange goes
beyond a purely single-active-electron model potential and provides a step
toward more \textit{ab initio} descriptions of Rydberg states. The resulting
comparisons help identify transitions whose \(E1\) matrix elements are robust
against, or intrinsically sensitive to, short-range modeling.

\section{Method}

\subsection{Dirac-Fock plus core-polarization method}

We write a one-electron radial Dirac spinor in the form
\begin{equation}
  \psi_{n\kappa m}(\mathbf{r}) =
  \frac{1}{r}
  \begin{pmatrix}
    P_{n\kappa}(r)\Omega_{\kappa m}(\hat{\mathbf{r}}) \\
    i Q_{n\kappa}(r)\Omega_{-\kappa m}(\hat{\mathbf{r}})
  \end{pmatrix},
\end{equation}
where $P_{n\kappa}$ and $Q_{n\kappa}$ are the large and small radial
components, respectively, $\Omega_{\kappa m}$ is the spinor spherical
harmonic, and $\kappa=(\ell-j)(2j+1)$. The closed-shell Rb$^+$ and Cs$^+$
ionic cores, with configurations [Kr] and [Xe], respectively, are obtained
using the DF method. For a core orbital $a$, the
radial core density used in the direct potential is
\begin{equation}
  \rho_c(r)=\sum_a (2j_a+1)\left[P_a^2(r)+Q_a^2(r)\right].
  \label{eq:core_density}
\end{equation}
The spherical Coulomb integral generated by this radial density is evaluated as
\begin{equation}
  Y_0(r)=\frac{1}{r}\int_0^r \rho_c(r')\,dr'
  +\int_r^\infty \frac{\rho_c(r')}{r'}\,dr',
  \label{eq:y0}
\end{equation}
and, with the sign convention used in the radial equations, the local direct
potential is
\begin{equation}
  V_{\mathrm{dir}}(r)=\frac{Z}{r}-Y_0(r).
  \label{eq:vdir}
\end{equation}
The radial equations for a valence or Rydberg orbital $v$ are then
\begin{align}
  \frac{dP_v}{dr} &=
  -\frac{\kappa_v}{r}P_v
  + \left(\frac{2}{\alpha}+\alpha[E_v+V_{\mathrm{loc}}(r)]\right)Q_v
  + \alpha X_{Q,v}(r),
  \label{eq:dirac_p}\\
  \frac{dQ_v}{dr} &=
  -\alpha[E_v+V_{\mathrm{loc}}(r)]P_v
  +\frac{\kappa_v}{r}Q_v
  -\alpha X_{P,v}(r),
  \label{eq:dirac_q}
\end{align}
where $V_{\mathrm{loc}}=V_{\mathrm{dir}}$ for a Rydberg orbital obtained with
the DF method, and $V_{\mathrm{loc}}=V_{\mathrm{dir}}+V_{\mathrm{CP}}$ for a
DFCP orbital. The nonlocal exchange vector
$X_v=(X_{P,v},X_{Q,v})^T$ is evaluated using the current valence orbital and
the frozen-core orbitals.

The same radial Coulomb-integral form is used for the exchange multipoles. For
two spinors $a$ and $b$, we define
\begin{equation}
  \rho_{ab}(r)=P_a(r)P_b(r)+Q_a(r)Q_b(r),
  \label{eq:transition_density}
\end{equation}
and
\begin{align}
  Y_k[ab](r) &=
  \frac{1}{r^{k+1}}\int_0^r r'^k\rho_{ab}(r')\,dr'
  + r^k\int_r^\infty\frac{\rho_{ab}(r')}{r'^{k+1}}\,dr' .
  \label{eq:yk}
\end{align}
For the closed-shell Rb and Cs cores, the nonlocal exchange term acting on the
Rydberg orbital is
\begin{equation}
  X_v(r)=
  \sum_c\sum_k
  C^k_{vc}Y_k[vc](r)
  \begin{pmatrix}P_c(r)\\Q_c(r)\end{pmatrix},
  \label{eq:exchange}
\end{equation}
where $c$ runs over occupied relativistic core subshells. The angular factor is
\begin{equation}
  C^k_{vc}=(2j_c+1)
  \left(
  \begin{array}{ccc}
  j_v & j_c & k\\
  1/2 & -1/2 & 0
  \end{array}
  \right)^2 .
  \label{eq:exchange_coeff}
\end{equation}
The allowed multipoles are
$k=|l_v-l_c|,|l_v-l_c|+2,\ldots,l_v+l_c$, which enforce the parity selection
of the exchange interaction.

Core-valence correlation is represented by the local core-polarization
potential \cite{PhysRevA.94.032503, PhysRevA.109.063115, cwhz-by5w,
PhysRevA.87.042517}
\begin{equation}
  V_{\mathrm{CP}}(r;\rho)=
  -\frac{\alpha_c}{2r^4}
  \left[
    1-\exp\left(-\frac{r^6}{\rho^6}\right)
  \right].
  \label{eq:vcp}
\end{equation}
We use the ionic-core $E1$ polarizabilities from the standard alkali
model-potential parametrization, $\alpha_c=9.076$ for Rb$^+$ and
$\alpha_c=15.644$ for Cs$^+$ \cite{PhysRevA.49.982}. The cutoff radius
$\rho$ regularizes the short-range behavior of the polarization potential and
is optimized for each $n\ell j$ state so that the DFCP energy reproduces a
reference energy obtained from the quantum-defect formula,
\begin{equation}
  E_n = -\frac{R}{[n-\delta(n)]^2}.
  \label{eq:qd_energy}
\end{equation}
Here $R$ is the atom-specific reduced-mass Rydberg constant used in the target
spectrum, with $R=0.49999676970706$ a.u. for Rb and $R=0.4999982667$ a.u. for
Cs; energies are measured relative to the ionization limit. The quantum defect
is represented as
\begin{equation}
  \delta(n) =
  \delta_0 + \frac{\delta_2}{(n-\delta_0)^2}
  + \frac{\delta_4}{(n-\delta_0)^4}.
  \label{eq:qd_expansion}
\end{equation}
When only two coefficients are available, the $\delta_4$ term is omitted.

For a trial value of $\rho$, the DFCP orbital is solved with exchange included. 
The Hellmann-Feynman derivative used in
the Newton iteration is
\begin{equation}
  \frac{dE}{d\rho}
  =
  \frac{3\alpha_c}{\rho^7}
  \int_0^\infty
  \left[P^2(r)+Q^2(r)\right]
  r^2\exp\left(-\frac{r^6}{\rho^6}\right)\,dr .
  \label{eq:rho_derivative}
\end{equation}
Equivalently, with
\begin{equation}
  I_\rho =
  \int_0^\infty
  \left[P^2(r)+Q^2(r)\right]
  r^2\exp\left(-\frac{r^6}{\rho^6}\right)\,dr ,
\end{equation}
the Newton update is
\begin{equation}
  \rho_{\mathrm{new}} =
  \rho +
  \frac{\left[E_{\mathrm{ref}}-E(\rho)\right]\rho^7}
       {3\alpha_c I_\rho}.
  \label{eq:rho_newton}
\end{equation}

\subsection{Adams-Moulton radial integration}

Rydberg orbitals are calculated on a logarithmic radial grid
$r_i=r_{\min}\exp[(i-1)h]$, with $r_{\min}=1.0\times10^{-6}$ a.u. and
$h=d\log r=0.003$. The grid is initially constructed with an outer boundary of
40 a.u. and is enlarged automatically when the orbital tail extends beyond
the current boundary. Direct Coulomb multipoles, core orbitals, and Rydberg
orbitals are represented on the same radial grid.

The integration is carried out in the logarithmic coordinate $x=\log r$,
for which the radial equations take the local linear form
\begin{equation}
  \frac{d\mathbf{y}}{dx}
  =
  r(x)\left[A(r)\mathbf{y}(x)+\mathbf{b}(r)\right],
  \qquad
  \mathbf{y}=(P,Q)^T .
  \label{eq:log_radial_equation}
\end{equation}
Here $\mathbf{b}$ contains the nonlocal exchange contribution when exchange is
included. For a trial energy, the outward integration is initialized using the
regular near-nucleus asymptotic form of the Dirac spinor. The inward integration
is initialized at the outer boundary using the exponentially decaying
bound-state asymptotic form. The amplitudes of the two starter solutions are
arbitrary, and their relative normalization is fixed at the matching point.

The 10th-order implicit Adams-Moulton step used for outward integration is
\begin{equation}
  \mathbf{y}_i =
  \mathbf{y}_{i-1}
  + \beta_0 h \mathbf{F}_i
  + \sum_{m=1}^{9}\beta_m h \mathbf{F}_{i-m},
  \label{eq:adams}
\end{equation}
where
\begin{equation}
  \mathbf{F}_i =
  r_i\left[A_i\mathbf{y}_i+\mathbf{b}_i\right],
  \label{eq:log_rhs}
\end{equation}
and $\beta_m$ are the fixed 10th-order Adams-Moulton coefficients. Because
$\mathbf{F}_i$ depends linearly on the unknown $\mathbf{y}_i$, each implicit
step reduces to a local $2\times2$ linear solve. The same formula is used for
inward integration with the step direction reversed. The bound-state energy is
iterated by matching the inward and outward solutions while enforcing the
required radial node count, which selects the desired Rydberg state directly.

For a given trial energy, the outer classical turning point is identified from
the current local potential as the point at which the bound-state motion changes
from classically allowed to forbidden. The matching point $r_m$ is chosen near
the maximum of the inward-integrated solution outside the inner core region.
After the outward and inward integrations are completed, the inward branch is
rescaled by $P_{\mathrm{out}}(r_m)/P_{\mathrm{in}}(r_m)$, with the same factor
applied to both spinor components. This removes the arbitrary tail
normalization and makes the large component continuous at $r_m$.

The energy is adjusted by requiring the logarithmic derivative of the large
component to be continuous at the matching point,
\begin{equation}
  F(E)=
  \frac{P'_{\mathrm{out}}(r_m)}{P_{\mathrm{out}}(r_m)}
  -
  \frac{P'_{\mathrm{in}}(r_m)}{P_{\mathrm{in}}(r_m)} ,
  \label{eq:matching_residual}
\end{equation}
where $P'$ is evaluated from the radial Dirac equation. The energy iteration is
continued until the matching residual and the energy change satisfy the chosen
convergence criteria.

The desired Rydberg state is selected by the radial node count of the large
component $P(r)$, with the near-nucleus region and the asymptotic tail excluded
from the count to avoid numerical sign changes. When exchange is included, the
nonlocal exchange source is recomputed after orbital updates. The orbitals are
renormalized after each update using Simpson quadrature on the logarithmic
radial grid. The same quadrature is used for normalization and radial matrix
elements,
\begin{equation}
  \int f(r)\,dr \simeq \sum_i w_i f(r_i),
  \label{eq:simpson_quadrature}
\end{equation}
where the weights $w_i$ are generated for the logarithmic grid.

\subsection{Rydberg-transition \texorpdfstring{$E1$}{E1} matrix elements}

All $E1$ matrix elements reported below are reduced matrix elements. The
angular factor used throughout the calculation is
\begin{equation}
  \begin{aligned}
  \langle \kappa_b||C^K||\kappa_a\rangle
  &=
  (-1)^\phi
  \sqrt{(2j_b+1)(2j_a+1)}
  \\
  &\quad\times
  \left(
  \begin{array}{ccc}
  j_b & j_a & K\\
  1/2 & -1/2 & 0
  \end{array}
  \right),
  \end{aligned}
  \label{eq:angular_coeff}
\end{equation}
where $\phi=t_b+\lfloor(t_a+1)/2\rfloor+K$, with
$t_i=2j_i=-2\kappa_i-1$ for $\kappa_i<0$ and
$t_i=2\kappa_i-1$ for $\kappa_i>0$. In the zero-frequency limit, the $E1$
length-form matrix element $D_L$ is
\begin{equation}
  D_L =
  \langle \kappa_b||C^1||\kappa_a\rangle
  \int_0^\infty
  \left[P_b(r)P_a(r)+Q_b(r)Q_a(r)\right]r\,dr .
  \label{eq:e1_length}
\end{equation}

The $E1$ velocity-form matrix element $D_V$ is obtained from the finite-frequency
operator and evaluated using its low-frequency expansion. For the microwave
transitions between Rydberg states considered here, this expansion is valid
over the total radial region,
where $pr\ll1$ and $p=|\omega|/c$. Defining
\begin{equation}
  \Delta\kappa=-\frac{\kappa_b-\kappa_a}{2},
\end{equation}
the Taylor coefficients entering the two spinor components are
\begin{align}
  c_g(r) &=
  \frac{2\Delta\kappa+1}{3}
  -\frac{(4\Delta\kappa+1)(pr)^2}{30}
  \notag\\
  &\quad
  +\frac{(6\Delta\kappa+1)(pr)^4}{840}
  -\frac{(8\Delta\kappa+1)(pr)^6}{45360},
  \label{eq:cg}\\
  c_f(r) &=
  \frac{2\Delta\kappa-1}{3}
  -\frac{(4\Delta\kappa-1)(pr)^2}{30}
  \notag\\
  &\quad
  +\frac{(6\Delta\kappa-1)(pr)^4}{840}
  -\frac{(8\Delta\kappa-1)(pr)^6}{45360}.
  \label{eq:cf}
\end{align}
The corresponding $E1$ velocity-form matrix element $D_V$ is
\begin{equation}
  \begin{aligned}
  D_V &=
  -\langle \kappa_b||C^1||\kappa_a\rangle
  \frac{3}{p}
  \\
  &\quad\times
  \int_0^\infty
  \left[P_b(r)c_g(r)Q_a(r)+Q_b(r)c_f(r)P_a(r)\right]\,dr .
  \end{aligned}
  \label{eq:e1_velocity}
\end{equation}
For the microwave transitions between Rydberg states considered here, terms
through $(pr)^6$ are retained in the low-frequency expansion. The residual
length-velocity difference reported below is therefore interpreted as an
internal gauge diagnostic rather than as an estimate of the
Taylor-truncation error. For the microwave-frequency Rydberg transitions, 
the conservative radial
scale $r\sim2n^2a_0$ gives $pr\lesssim3.2\times10^{-3}$, so terms beyond
$(pr)^6$ are negligible on the scale of the reported length-velocity
differences.

In addition to the ordinary length-form matrix element, we also evaluate
the CP-corrected length-form matrix element, denoted $D_{\rm CP}$, using the
effective dipole operator associated with the core-polarization potential
\cite{PhysRevA.49.982, PhysRevA.87.042517},
\begin{equation}
  D_{\mathrm{eff}}(r)
  =
  r
  \left[
    1
    -
    \frac{\alpha_c}{r^3}
    \sqrt{
      1-\exp\left(-\frac{r^6}{\rho^6}\right)
    }
  \right].
  \label{eq:deff}
\end{equation}
For a transition
between two DFCP orbitals with cutoff radii $\rho_b$ and $\rho_a$, we use the
mean value $\rho = \frac{\rho_b+\rho_a}{2}$ in the effective operator.

Equation~\eqref{eq:deff} is the cutoff-regularized
form of the long-range static local limit of the random-phase approximation
(RPA) core-response correction, as derived in
Appendix~\ref{app:cp_length_rpa}. We use it to estimate the operator-level
core-polarization correction. A previous DFCP study of
Th$^{3+}$ found that this effective operator has a noticeable effect on
calculated matrix elements \cite{PhysRevA.109.063115}. 

The core-polarization potential can be viewed as a long-range local
approximation to the second-order correlation self-energy
operator \cite{Johnson2007AtomicStructureTheory}, while the CP-corrected
length-form operator represents the corresponding long-range static local
approximation to the RPA core response. Thus, DFCP matrix elements evaluated
with this effective operator may be regarded as a reliable approximation to
the dominant contributions included in a second-order
MBPT treatment combined with RPA corrections to the $E1$ amplitude.

\subsection{Hyperfine-structure constants}

Hyperfine-structure (HFS) constants are used to test the near-core behavior of
the calculated Rydberg orbitals. The reduced electronic HFS matrix elements
are evaluated using the same angular convention as Eq.~\eqref{eq:angular_coeff},
together with the finite-nuclear-size radial factor
\begin{equation}
  F_K(r)=
  \begin{cases}
  r/R_N^{K+2}, & r<R_N,\\
  r^{-(K+1)}, & r\ge R_N,
  \end{cases}
  \label{eq:hfs_finite_nucleus}
\end{equation}
where $R_N$ is the nuclear radius. The reduced
matrix element of the magnetic rank-$K$ hyperfine operator is
\begin{equation}
  \begin{aligned}
  \langle b||T^M_K||a\rangle &=
  -\frac{\kappa_a+\kappa_b}{K}
  \langle-\kappa_b||C^K||\kappa_a\rangle
  \\
  &\quad\times
  \int_0^\infty
  \left[P_b(r)Q_a(r)+Q_b(r)P_a(r)\right]F_K(r)\,dr ,
  \end{aligned}
  \label{eq:hfs_magnetic}
\end{equation}
and that of the electric rank-$K$ hyperfine operator is
\begin{equation}
  \begin{aligned}
  \langle b||T^E_K||a\rangle &=
  -\langle\kappa_b||C^K||\kappa_a\rangle
  \\
  &\quad\times
  \int_0^\infty
  \left[P_b(r)P_a(r)+Q_b(r)Q_a(r)\right]F_K(r)\,dr .
  \end{aligned}
  \label{eq:hfs_electric}
\end{equation}

For a diagonal matrix element in a state with angular momentum $J$, the
magnetic-dipole HFS constant $A$ in MHz is
\begin{equation}
  A =
  C_A\frac{\mu/I}{\sqrt{J(J+1)(2J+1)}}
  \langle J||T^M_1||J\rangle ,
  \label{eq:hfs_a}
\end{equation}
where $C_A=(\alpha/2)(m_e/m_p)(E_h/h)$ converts the electronic matrix element
in atomic units to MHz for $\mu$ expressed in nuclear magnetons. The
electric-quadrupole HFS constant $B$ is
\begin{equation}
  B =
  C_B Q
  \,2\sqrt{\frac{J(2J-1)}
  {(J+1)(2J+1)(2J+3)}}
  \langle J||T^E_2||J\rangle ,
  \label{eq:hfs_b}
\end{equation}
where $C_B=(1~\mathrm{barn}/a_0^2)(E_h/h)$ in MHz and $Q$ is expressed in
barns. The nuclear inputs are $I=5/2$, $\mu=1.3533515\,\mu_N$, and
$Q=0.276$ b for $^{85}$Rb, and $I=7/2$, $\mu=2.5829128\,\mu_N$, and
$Q=-0.00343$ b for $^{133}$Cs \cite{stoneTableNuclearMagnetic2005}.

\section{Results}

\subsection{Cutoff radii and energy validation}

Table~\ref{tab:qd_params} lists the quantum-defect parameters used to define
the target energies. These energies determine the DFCP cutoff radii and thereby
allow the calculated radial DFCP wavefunctions.

\begin{table*}
\caption{\label{tab:qd_params}Quantum-defect parameters used to define the target energies. The
expansion is given in Eq.~\eqref{eq:qd_expansion}.}
\begin{ruledtabular}
\begin{tabular}{lccccp{0.28\textwidth}}
Atom & $lj$ & $\delta_0$ & $\delta_2$ & $\delta_4$ & References \\
\hline
Rb & $s_{1/2}$ & 3.131 180 7(8) & 0.178 7(2) & 0.04(48) & Mack et al. \cite{PhysRevA.83.052515}; $\delta_4$: True et al. \cite{5wyj-9jqj} \\
Rb & $p_{1/2}$ & 2.654 746(45) & 0.290 & & Li et al. \cite{PhysRevA.99.042502} \\
Rb & $p_{3/2}$ & 2.641 657(20) & 0.295 & & Li et al. \cite{PhysRevA.99.042502} \\
Rb & $d_{3/2}$ & 1.348 094 8(11) & $-$0.605 4(4) & $-$0.77(31) & Mack et al. \cite{PhysRevA.83.052515}; $\delta_4$: True et al. \cite{5wyj-9jqj} \\
Rb & $d_{5/2}$ & 1.346 462 2(11) & $-$0.594 0(4) & $-$0.80(29) & Mack et al. \cite{PhysRevA.83.052515}; $\delta_4$: True et al. \cite{5wyj-9jqj} \\
Rb & $f_{5/2}$ & 0.016 519 2(9) & $-$0.085(9) & & Han et al. \cite{PhysRevA.74.054502} \\
Rb & $f_{7/2}$ & 0.016 543 7(7) & $-$0.086(7) & & Han et al. \cite{PhysRevA.74.054502} \\
Cs & $s_{1/2}$ & 4.049 359 94(17) & 0.238 017(18) & 0.174 7(33) & Shen et al. \cite{PhysRevLett.133.233005} \\
Cs & $p_{1/2}$ & 3.591 587 1(3) & 0.362 73(16) & & Deiglmayr et al. \cite{PhysRevA.93.013424} \\
Cs & $p_{3/2}$ & 3.559 067 6(3) & 0.374 69(14) & & Deiglmayr et al. \cite{PhysRevA.93.013424} \\
Cs & $d_{3/2}$ & 2.475 458 40(13) & 0.008 339(62) & $-$0.376 9(95) & Shen et al. \cite{PhysRevLett.133.233005} \\
Cs & $d_{5/2}$ & 2.466 315 29(14) & 0.013 431(65) & $-$0.361 3(61) & Shen et al. \cite{PhysRevLett.133.233005} \\
Cs & $f_{5/2}$ & 0.033 414 93(18) & $-$0.200 36(14) & 0.282 5(8) & Shen et al. \cite{dkb4-kb5b} \\
Cs & $f_{7/2}$ & 0.033 562 89(19) & $-$0.202 89(14) & 0.299 8(9) & Shen et al. \cite{dkb4-kb5b} \\
\end{tabular}
\end{ruledtabular}
\end{table*}

The first step is to determine a separate cutoff radius for each $n\ell j$
state using the Newton procedure in Eq.~\eqref{eq:rho_newton}. This inversion
becomes ill-conditioned at high $n$: the Rydberg electron spends most of its
time far from the core, so the energy derivative with respect to $\rho$
becomes small. Very small energy shifts can then produce large changes in the
inferred cutoff radius. The fit is therefore restricted to the
lower but still Rydberg-like range $15\le n\le 40$, where the inversion is
relatively stable. The resulting cutoff radii are fitted with the linear-exponential
model
\begin{equation}
  \rho_{\ell j}(n)=\rho_0+A n+B\exp(-n/\tau),
  \label{eq:rho_fit}
\end{equation}
and the fitted $\rho_{\ell j}(n)$ is tested by recalculating the DFCP energies
over the target Rydberg range.

\begin{table*}
\caption{\label{tab:rho_fit_params}Parameters of the linear-exponential
cutoff-radius fit in Eq.~\eqref{eq:rho_fit}. Residuals are given in atomic
units of length.}
\begin{ruledtabular}
\begin{tabular}{llrrrrrr}
Atom & $lj$ & $\rho_0$ & $A$ & $B$ & $\tau$ & RMS & Max. abs. \\
\hline
Rb & $s_{1/2}$ & 2.338099 & \num{8.088e-5} & 0.025285 & 5 & \num{7.537e-6} & \num{2.420e-5} \\
Rb & $p_{1/2}$ & 2.270244 & \num{1.068e-4} & 0.080209 & 4 & \num{1.460e-5} & \num{3.195e-5} \\
Rb & $p_{3/2}$ & 2.267477 & \num{1.046e-4} & 0.078319 & 4 & \num{1.420e-5} & \num{3.110e-5} \\
Rb & $d_{3/2}$ & 2.719788 & \num{2.505e-6} & 0.030684 & 6 & \num{7.894e-6} & \num{2.452e-5} \\
Rb & $d_{5/2}$ & 2.735318 & \num{2.716e-6} & 0.031640 & 6 & \num{8.328e-6} & \num{2.600e-5} \\
Rb & $f_{5/2}$ & 2.595970 & \num{3.260e-3} & 0.326016 & 5 & \num{5.188e-5} & \num{1.104e-4} \\
Rb & $f_{7/2}$ & 2.613224 & \num{3.198e-3} & 0.370800 & 5 & \num{6.477e-5} & \num{1.431e-4} \\
Cs & $s_{1/2}$ & 2.657160 & \num{2.237e-5} & 0.077455 & 4 & \num{1.395e-5} & \num{3.038e-5} \\
Cs & $p_{1/2}$ & 2.545430 & \num{2.823e-5} & 0.113232 & 4 & \num{1.493e-5} & \num{2.906e-5} \\
Cs & $p_{3/2}$ & 2.568989 & \num{2.868e-5} & 0.111078 & 4 & \num{1.457e-5} & \num{2.799e-5} \\
Cs & $d_{3/2}$ & 3.057658 & \num{-2.428e-5} & 0.073552 & 5 & \num{1.343e-5} & \num{2.853e-5} \\
Cs & $d_{5/2}$ & 3.093270 & \num{-2.608e-5} & 0.078292 & 5 & \num{1.432e-5} & \num{3.051e-5} \\
Cs & $f_{5/2}$ & 2.703158 & \num{4.494e-4} & 0.010097 & 6 & \num{3.181e-6} & \num{7.762e-6} \\
Cs & $f_{7/2}$ & 2.737737 & \num{4.427e-4} & 0.020428 & 6 & \num{4.552e-6} & \num{1.069e-5} \\
\end{tabular}
\end{ruledtabular}
\end{table*}

\begin{figure*}
  \includegraphics[width=\textwidth]{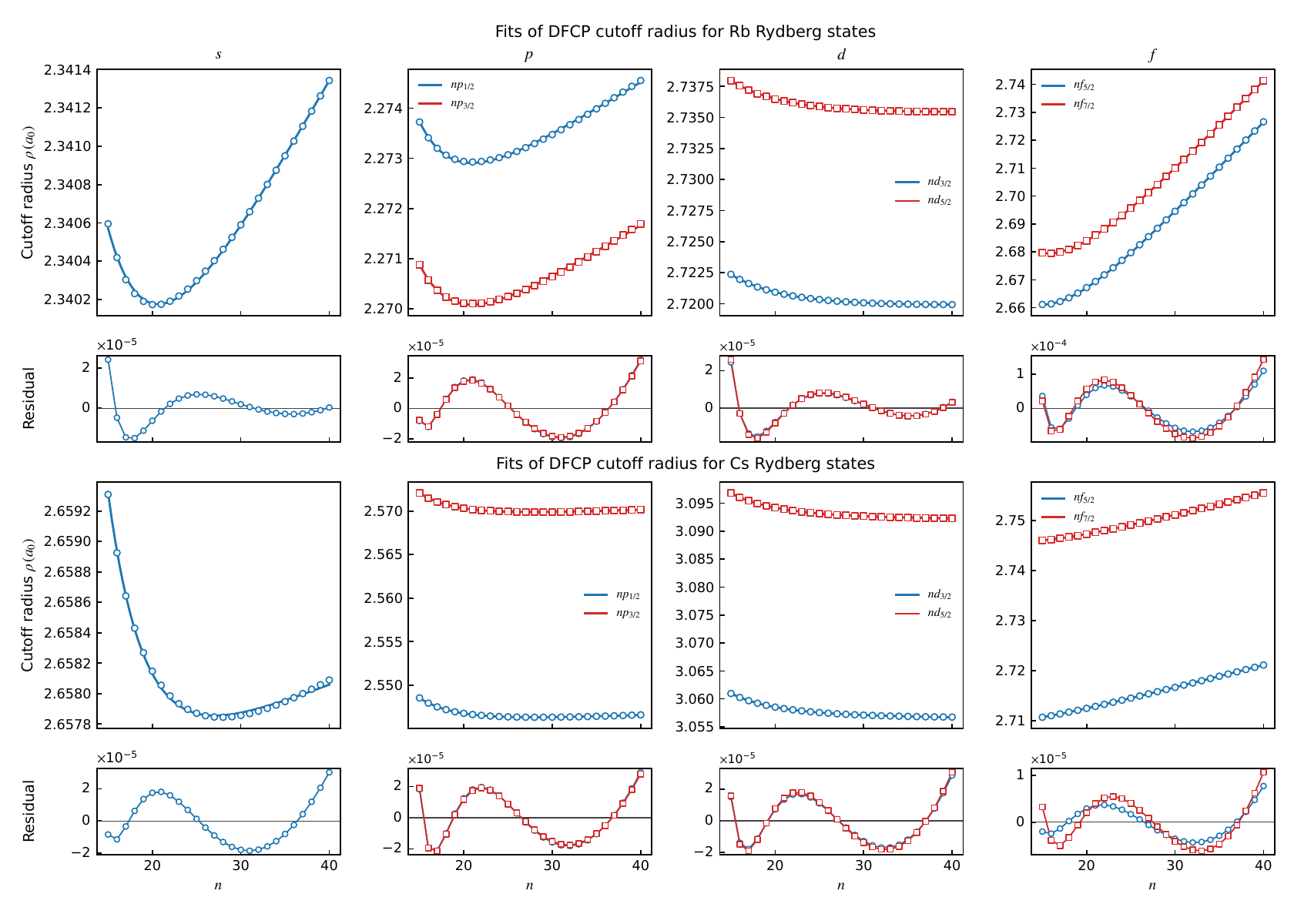}
  \caption{\label{fig:rho_fit}Fitted DFCP cutoff radii $\rho(n)$ for the Rb and
Cs $s$, $p$, $d$, and $f$ series. The fit uses cutoff radii determined
individually for states in the stable range $15\le n\le 40$ and provides the
smooth cutoff-radius model used for the higher-Rydberg orbitals.}
\end{figure*}

Figure~\ref{fig:rho_fit} and Table~\ref{tab:rho_fit_params} show the fitted
cutoff radii.  The $f$
series are the least constrained because the corresponding orbitals have very
small quantum defects and weak core penetration. Nevertheless, the fitted
residuals remain below $1.5\times10^{-4}$ a.u. for all retained Rb and Cs
series.

The integer values of $\tau$ in Table~\ref{tab:rho_fit_params} result from the
fitting procedure. Several fixed values of $\tau$ were scanned; for each value,
the remaining parameters $\rho_0$, $A$, and $B$ were obtained by linear least
squares, and the fit with the smallest RMS residual was retained. This
discrete scan does not affect the conclusions, since the Rydberg wavefunctions
and Rydberg-transition $E1$ matrix elements are weakly sensitive to changes in
$\rho$ on the scale of the fit residuals, as verified by the cutoff-radius
perturbation test discussed below.

\begin{table}
\caption{\label{tab:energy_validation}Energy residuals for the fitted
$\rho(n)$ model. Residuals are DFCP energies minus quantum-defect target
energies, expressed in MHz. The fit region is $15\le n\le 40$, and the
extrapolation region is $41\le n\le 90$.}
\begin{ruledtabular}
\begin{tabular}{lccc}
Atom & Region & RMS & Max. abs. \\
\hline
Rb & Fit & 1.25 & 9.49 \\
Rb & Extrapolation & 0.83 & 1.41 \\
Cs & Fit & 1.28 & 6.68 \\
Cs & Extrapolation & 1.40 & 2.67 \\
\end{tabular}
\end{ruledtabular}
\end{table}

\begin{figure*}
  \includegraphics[width=\textwidth]{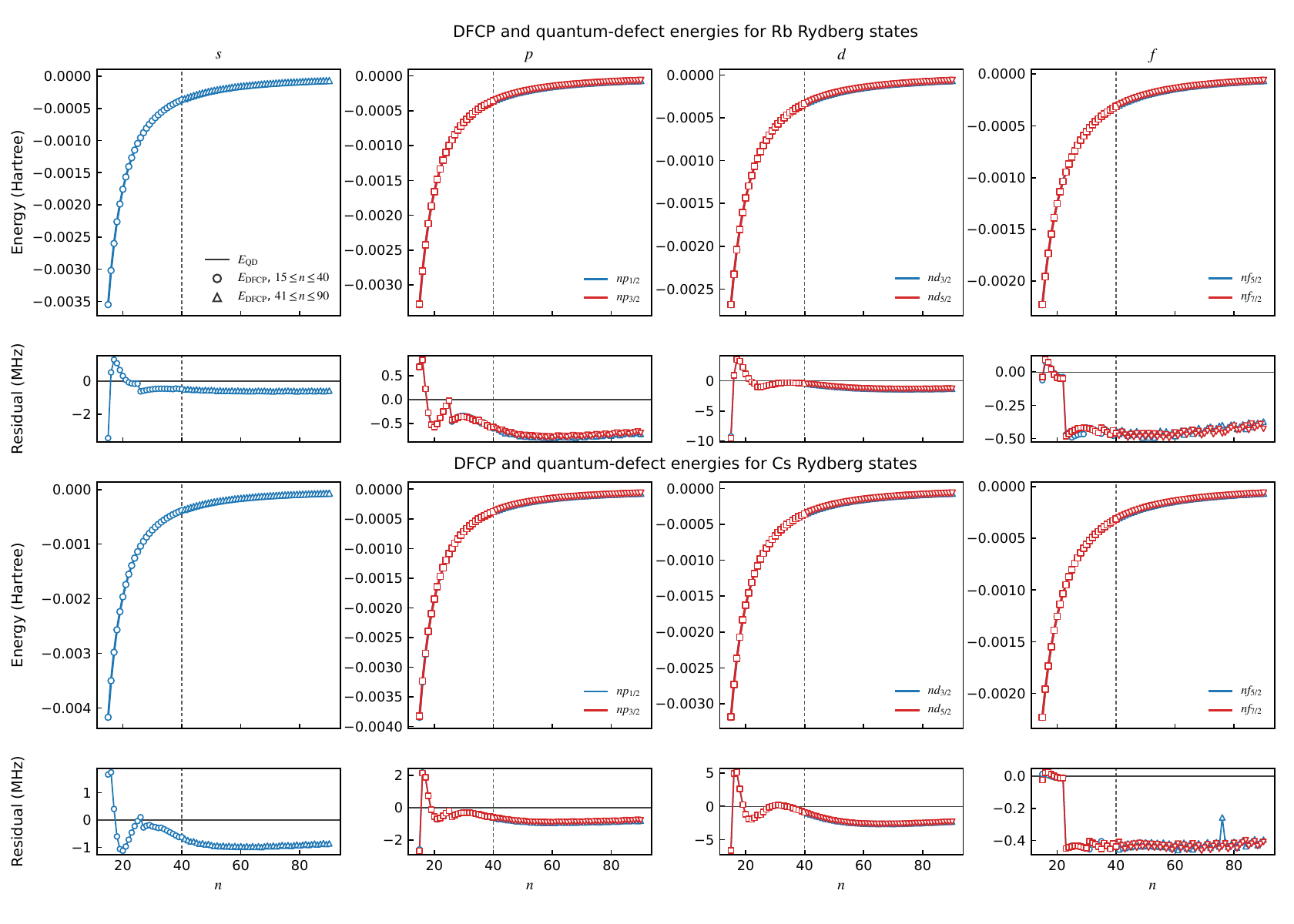}
  \caption{\label{fig:energy_validation}DFCP energy residuals obtained with the
fitted cutoff-radius model. The comparison extends to $n=90$, beyond the
fitting interval used to determine $\rho(n)$.}
\end{figure*}

The energy residuals in Fig.~\ref{fig:energy_validation} and
Table~\ref{tab:energy_validation} show that the fitted cutoff-radius model
continues to reproduce the quantum-defect target energies beyond the fitting
range. In the extrapolation region, the RMS residuals are 0.83 MHz for Rb and
1.40 MHz for Cs, with maximum absolute residuals of 1.41 MHz and 2.67 MHz,
respectively. These residuals are small on the scale relevant for the present
Rydberg-orbital calculations. 

The smooth $\rho(n)$ model avoids amplifying small high-$n$ energy residuals
into unstable variations of the cutoff radius. It provides a DFCP potential
that reproduces the target energies while generating consistent Rydberg
orbitals over the full range considered here.

\subsection{Wavefunction diagnostics}

The logarithmic radial grid must represent both compact core orbitals and
extended Rydberg orbitals. This also allows the Rydberg spinors to be resolved
near the nucleus, where HFS constants are most sensitive.
Because these constants sample the short-range wavefunction more strongly than
Rydberg-transition $E1$ matrix elements, they provide a diagnostic of the
near-core behavior generated by three Hamiltonians: the direct-potential-only
Hamiltonian (hereafter direct), the DF Hamiltonian, and the DFCP Hamiltonian.
Table~\ref{tab:hfs} compares the calculated scaled constants,
$A_{\rm sc}=A[n-\delta(n)]^3$, with available measurements, where $\delta(n)$
is the quantum-defect function in Eq.~\eqref{eq:qd_expansion}.

\begin{table}
\caption{\label{tab:hfs}Scaled HFS magnetic-dipole constants
$A[n-\delta(n)]^3$ in MHz for $^{85}$Rb and $^{133}$Cs.}
\begin{ruledtabular}
\begin{tabular}{llrl}
Atom & State & DFCP & Expt. \\
\hline
Rb & $s_{1/2}$ & 4605 & $5124(26)$ \cite{PhysRevA.100.062515} \\
Rb & $p_{1/2}$ & 1125 & $1443(31)$ \cite{PhysRevA.106.052810} \\
Rb & $p_{3/2}$ & 201 & -- \\
Rb & $d_{3/2}$ & 185 & -- \\
Rb & $d_{5/2}$ & 77 & -- \\
Rb & $f_{5/2}$ & 2 & -- \\
Rb & $f_{7/2}$ & 1 & -- \\
Cs & $s_{1/2}$ & 11510 & -- \\
Cs & $p_{1/2}$ & 2852 & $3760(26)$ \cite{PhysRevA.110.042815} \\
Cs & $p_{3/2}$ & 430 & $718(27)$ \cite{PhysRevA.110.042815} \\
Cs & $d_{3/2}$ & 416 & -- \\
Cs & $d_{5/2}$ & 168 & -- \\
Cs & $f_{5/2}$ & 3 & -- \\
Cs & $f_{7/2}$ & 2 & -- \\
\end{tabular}
\end{ruledtabular}
\end{table}


Penetrating states show substantial Hamiltonian dependence. For Rb
$p_{1/2}$, for example, the scaled constant changes from 2533 MHz in the
direct calculation to 983 MHz in DF and 1125 MHz in DFCP. The Rb $d_{3/2}$
value changes in the opposite direction, from 29 MHz to 145 MHz and then to
185 MHz. This sensitivity is expected because HFS constants depend on the
near-core wavefunction, where exchange and core-polarization effects are
largest. The $f$ states are much less affected because they penetrate the
ionic core weakly.

Comparison with experiment gives the scale of the near-core accuracy. The
DFCP values are 4605 MHz for Rb $s_{1/2}$ compared with $5124(26)$ MHz, and
1125 MHz for Rb $p_{1/2}$ compared with $1443(31)$ MHz. For Cs, the DFCP
values are 2852 MHz for $p_{1/2}$ compared with $3760(26)$ MHz, and
430 MHz for $p_{3/2}$ compared with $718(27)$ MHz. The agreement is therefore
at the level of tens of percent for the measured low-angular-momentum HFS
constants. This should not be interpreted as a quantitative validation of the
wavefunction. Rather, it shows that the DFCP orbitals have the
correct order of magnitude and reasonable scaling for short-range observables,
while also indicating the expected limitation of the present treatment in the
core region. The DFCP Hamiltonian is optimized to reproduce Rydberg energies,
not HFS constants. In particular, the core-polarization potential is a
long-range approximation and does not represent the short-range correlations 
that dominate high-accuracy HFS
calculations.

\begin{table}
\caption{\label{tab:low_lying_e1_allorder}
Comparison of low-lying reduced $E1$ matrix elements with all-order reference
values. $D_L$, $D_V$, and $D_{\rm CP}$ are calculated with the DFCP
wavefunctions using the length form, the velocity form, and the CP-corrected
length-form operator, respectively. All entries are absolute values in atomic
units.}
\begin{ruledtabular}
\begin{tabular}{lrrrr}
Transition & $D_L$ & $D_V$ & $D_{\rm CP}$ & Ref. \cite{PhysRevA.69.022509, PhysRevA.94.012505} \\
\hline
\multicolumn{5}{c}{Rb} \\
\hline
$5s_{1/2}\to5p_{1/2}$ & 4.471 & 4.287 & 4.218 & 4.221 \\
$5s_{1/2}\to5p_{3/2}$ & 6.305 & 6.037 & 5.958 & 5.956 \\
$5p_{1/2}\to6d_{3/2}$ & 1.025 & 0.905 & 1.087 & 1.180 \\
$5p_{3/2}\to6d_{3/2}$ & 0.489 & 0.433 & 0.515 & 0.558 \\
$5p_{3/2}\to6d_{5/2}$ & 1.459 & 1.290 & 1.537 & 1.658 \\
$6d_{3/2}\to4f_{5/2}$ & 9.281 & 9.329 & 9.286 & 9.938 \\
$6d_{5/2}\to4f_{5/2}$ & 2.474 & 2.487 & 2.475 & 2.642 \\
$6d_{5/2}\to4f_{7/2}$ & 11.06 & 11.12 & 11.07 & 11.81 \\
\hline
\multicolumn{5}{c}{Cs} \\
\hline
$7s_{1/2}\to7p_{1/2}$ & 10.39 & 10.18 & 10.29 & 10.31 \\
$7s_{1/2}\to7p_{3/2}$ & 14.43 & 14.12 & 14.30 & 14.32 \\
$7p_{1/2}\to6d_{3/2}$ & 18.13 & 20.51 & 18.01 & 17.99 \\
$7p_{3/2}\to6d_{3/2}$ & 8.14 & 9.57 & 8.08 & 8.07 \\
$7p_{3/2}\to6d_{5/2}$ & 24.54 & 28.71 & 24.39 & 24.35 \\
$6d_{3/2}\to4f_{5/2}$ & 24.72 & 24.83 & 24.65 & 24.62 \\
$6d_{5/2}\to4f_{5/2}$ & 6.63 & 6.66 & 6.61 & 6.60 \\
$6d_{5/2}\to4f_{7/2}$ & 29.64 & 29.79 & 29.56 & 29.50 \\
\end{tabular}
\end{ruledtabular}
\end{table}

The HFS comparison therefore provides a stringent near-core test, but HFS
accuracy is not equivalent to the accuracy of $E1$ matrix elements.
To provide a more direct diagnostics of Rydberg-transition $E1$ matrix elements, 
Table~\ref{tab:low_lying_e1_allorder}
compares low-lying $E1$ matrix elements calculated with the DFCP
wavefunctions against high-accuracy all-order reference values. Compared 
with $D_L$, the CP-corrected length-form operator substantially
improves the agreement between the DFCP and all-order results. Except 
for transitions involving Rb $d$ and
$f$ states, the relative differences between $D_{\rm CP}$ and the all-order
values are below 0.3\%. This comparison indicates that, although HFS constants
expose the limitations of the short-range wavefunction, the DFCP
wavefunctions combined with the CP-corrected length-form operator can still provide
accurate $E1$ matrix elements for representative low-lying
transitions.

This distinction is consistent with previous DFCP experience. For alkali
atoms, a DFCP calculation of ground-state magnetic-dipole HFS constants found
deviations at approximately the 10\% level \cite{cwhz-by5w}, comparable in
scale to the present Rydberg-state results. In contrast, a previous DFCP
study of the strongly core-valence-correlated Th$^{3+}$ system showed much
larger discrepancies between calculated and measured hyperfine constants,
while the $E1$ matrix elements still agreed with all-order results
at roughly the 10\% level or better \cite{PhysRevA.109.063115}. These
comparisons support the interpretation that HFS constants provide a stringent
short-range diagnostic, whereas $E1$ matrix elements are generally
less dominated by the near-nuclear wavefunction.

\subsection{Rydberg-transition \texorpdfstring{$E1$}{E1} matrix elements}

\subsubsection{Hamiltonian dependence and radial-integral cancellation}

As a test of the sensitivity to the fitted cutoff radii, we sampled
240 Rb and Cs transitions from the range $30\le n\le70$ with
$n'=n+\Delta n$ and $\Delta n\in\{-3,-2,-1,0,1,2,3\}$. The sample included the
$ns_{1/2}\rightarrow n'p_{1/2}$,
$np_{3/2}\rightarrow n'd_{5/2}$, and
$nd_{5/2}\rightarrow n'f_{7/2}$ branches. For each transition, the two DFCP
orbitals were recalculated after independently perturbing the two cutoff
radii by
\(\delta\rho_a,\delta\rho_b=\pm5\times10^{-4}\) a.u., with independently
random signs. The relative change in the length-form
matrix element was then recorded as
\begin{equation}
  \Delta_\rho =
  \frac{\bigl||D_L(\rho_a+\delta\rho_a,\rho_b+\delta\rho_b)|
  -|D_L(\rho_a,\rho_b)|\bigr|}
       {|D_L(\rho_a,\rho_b)|}.
  \label{eq:rho_sensitivity}
\end{equation}
The perturbation amplitude is deliberately conservative: the largest
cutoff-radius residual in Table~\ref{tab:rho_fit_params} is
$1.43\times10^{-4}$ a.u., well below $5\times10^{-4}$ a.u.

This perturbation test shows that variations of the fitted cutoff radii on
the scale of the fit residuals have limited effects on large matrix elements,
whereas strongly cancelled transitions can show large relative responses.
Over the full random sample, the median value of $\Delta_\rho$ is
$1.5\times10^{-4}$ and the 90th percentile is $4.0\times10^{-3}$. The largest
response occurs for the Rb $43d_{5/2}\rightarrow45f_{7/2}$ transition, for
which the baseline matrix element is only $|D_L|=0.115$ a.u. and the
perturbation changes it to $|D_L|=0.165$ a.u., giving
$\Delta_\rho=0.431$. This large relative change reflects cancellation in a
small residual matrix element and is not an indication of comparable
uncertainty for large matrix elements. For cases with $|D_L|>100$ a.u., the
largest relative change in the sample is $1.6\times10^{-3}$; for the robust
$s$-$p$ branches, it remains below $1.9\times10^{-4}$.


We next compare Rydberg-transition $E1$ matrix elements computed with three
Hamiltonians: direct, DF, and DFCP. The direct calculation includes the nuclear
and direct core Coulomb potentials. The DF calculation adds exchange with the
closed-shell core, and the DFCP calculation further includes the
core-polarization potential. The direct--DF and DF--DFCP differences therefore
separate the orbital-level effects of exchange and core polarization on the
same radial integral.

\begin{figure*}
  \includegraphics[width=\textwidth]{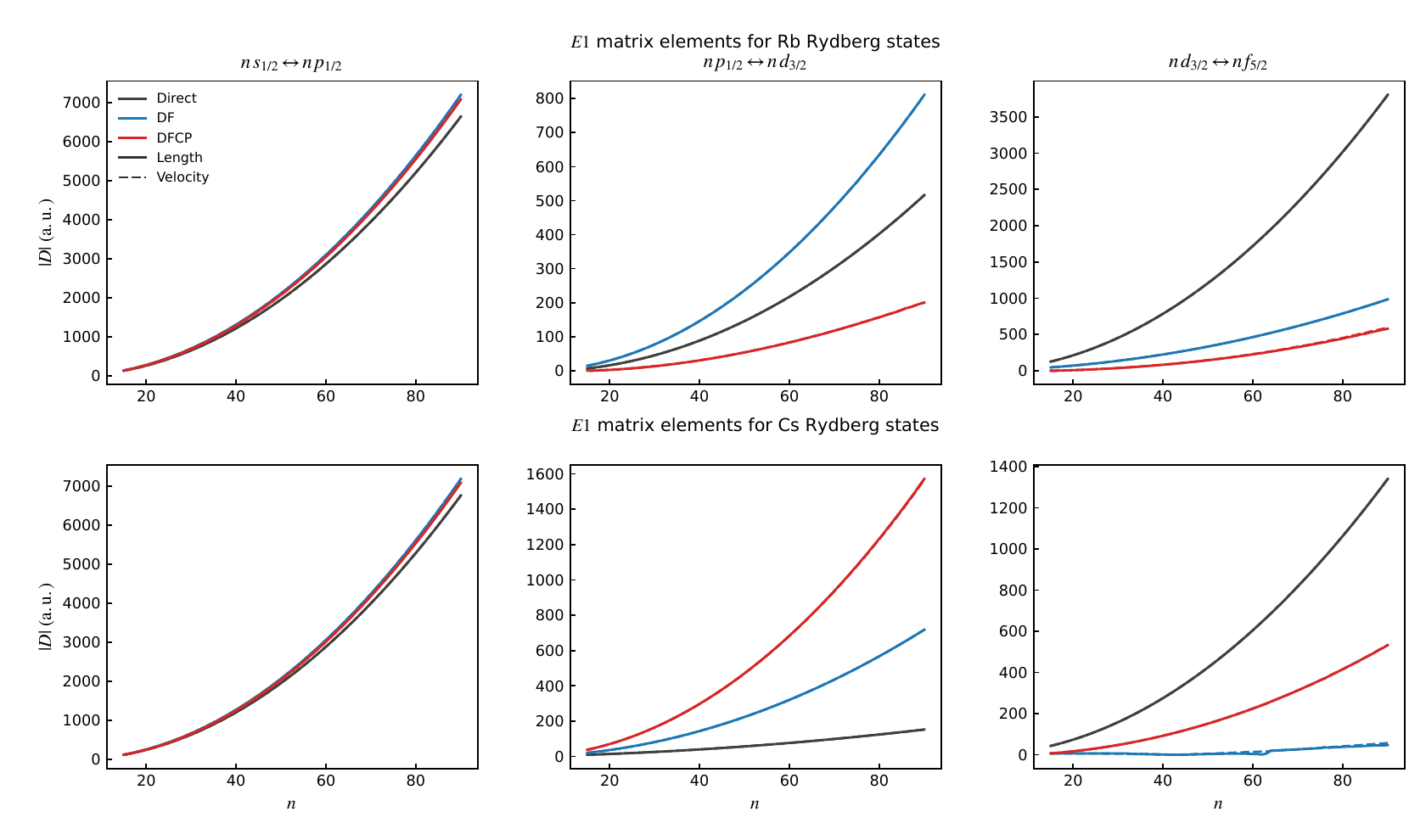}
  \caption{\label{fig:same_n_layers}Same-$n$ Rydberg-transition $E1$ matrix
  elements computed with direct, DF, and DFCP orbitals. The comparison shows
  orbital-level changes caused by explicit core exchange and by the additional
  core-polarization potential.}
\end{figure*}

Figure~\ref{fig:same_n_layers} shows that the largest Hamiltonian-dependent
changes occur in the $p$-$d$ and $d$-$f$ transitions. The direct--DF
differences reflect the orbital-level effect of explicit exchange with the
closed-shell core, while the DF--DFCP differences reflect the additional
effect of the core-polarization potential. Large changes across these
Hamiltonians indicate that the corresponding Rydberg-transition $E1$ matrix
element is not determined solely by hydrogenic outer-region potential, but
remains sensitive to the short-range effective potential.

This sensitivity can be understood from the radial integrand rather than from
the energy shift alone. Exchange and core polarization act mainly at short
range, but the associated short-range phase shift changes the node positions
and radial phase throughout the outer Rydberg region. When the cumulative
$E1$ integral is dominated by contributions of the same sign and saturates
smoothly, such changes have a limited effect on the final value. When the
cumulative integral contains large positive and negative contributions, the
final matrix element is a small difference between larger terms and can be
strongly affected by a small phase or amplitude change.

\begin{figure*}
  \includegraphics[width=\textwidth]{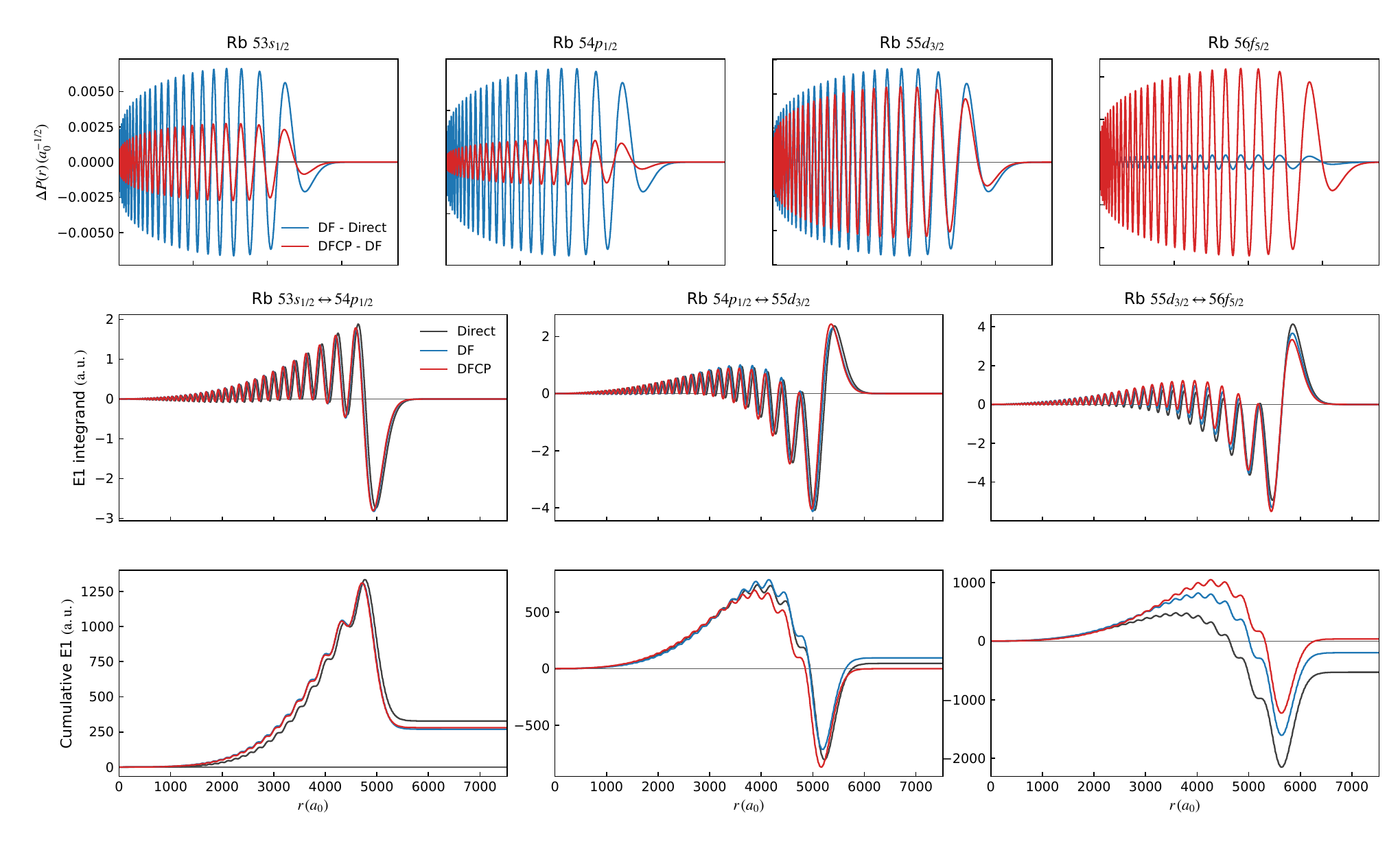}
  \caption{\label{fig:integrand_diagnostics}Rb cross-$n$ comparison for
  representative $s$-$p$, $p$-$d$, and $d$-$f$ transitions. The panels compare
  wavefunction differences, radial integrands, and cumulative radial
  integrals for the direct, DF, and DFCP Hamiltonians.}
\end{figure*}

To display this mechanism directly, we select representative examples shown in
Fig.~\ref{fig:integrand_diagnostics}. The
$s$-$p$ example has a comparatively smooth cumulative integral, while the
$p$-$d$ and $d$-$f$ examples show stronger cancellation between positive and
negative contributions. Because the two orbitals in these transitions have
different degrees of core penetration, exchange and core polarization change
their node positions, amplitudes, and radial phases differently. The resulting
noncoherent orbital changes make the cancelled $E1$ matrix elements more
sensitive to the short-range Hamiltonian.

\subsubsection{Model sensitivity and reliability classification}

We compare the present DFCP reduced Rydberg-transition $E1$ matrix elements
with model-potential values for Rb and Cs over $20\le n\le 89$ and
$\Delta n=-1,0,+1$. These reference values are obtained from the ARC package
\cite{SibalicARCOpensourceLibrary2017,robertsonARC30Expanded2021}, which
implements the standard alkali Rydberg model-potential calculation used for
these matrix elements. The relative difference from the model-potential value
is defined as
\begin{equation}
  \Delta_{\rm model}=
  \frac{\bigl||D_L|-|D_{\mathrm{model}}|\bigr|}
       {|D_L|}.
  \label{eq:model_difference}
\end{equation}
The internal length-velocity difference is defined as
\begin{equation}
  \Delta_{V-L}=
  \frac{\bigl||D_V|-|D_L|\bigr|}{|D_L|}.
  \label{eq:internal_difference}
\end{equation}
The purpose of this comparison is to assess model sensitivity rather than to
assign an absolute uncertainty to the calculated matrix elements. It identifies
branches whose matrix elements are insensitive to the short-range description,
and branches that show sizeable differences between the present DFCP orbitals
and the model-potential orbitals.

The combined evidence provides an operational reliability classification rather
than an absolute uncertainty estimate. We use $D_{\min}=100~\mathrm{a.u.}$
as a practical scale for identifying matrix elements whose relative
differences may be amplified by radial-integral cancellation. This threshold
is not a sharp physical boundary; it separates large Rydberg-transition $E1$
matrix elements from smaller branches for which a small residual matrix
element can make relative differences misleadingly large. The 1\% thresholds
for $\Delta_{\rm model}$ and $\Delta_{V-L}$ are likewise used only as
practical criteria for organizing the data.

With these criteria, a robust matrix element satisfies
$\Delta_{\rm model}<1\%$, $\Delta_{V-L}<1\%$, and $|D_L|>D_{\min}$. A
model-sensitive matrix element has $\Delta_{\rm model}>1\%$ while not being
dominated by the small-denominator condition. A gauge-sensitive matrix element
has $\Delta_{V-L}>1\%$. A cancellation-sensitive matrix element has
$|D_L|<D_{\min}$, a strong cancellation between positive and negative
contributions in the cumulative radial integral, or a nearby zero crossing.
These labels are diagnostic and can overlap, since radial-integral
cancellation can amplify both model differences and length-velocity
differences.

\begin{figure*}
  \includegraphics[width=\textwidth]{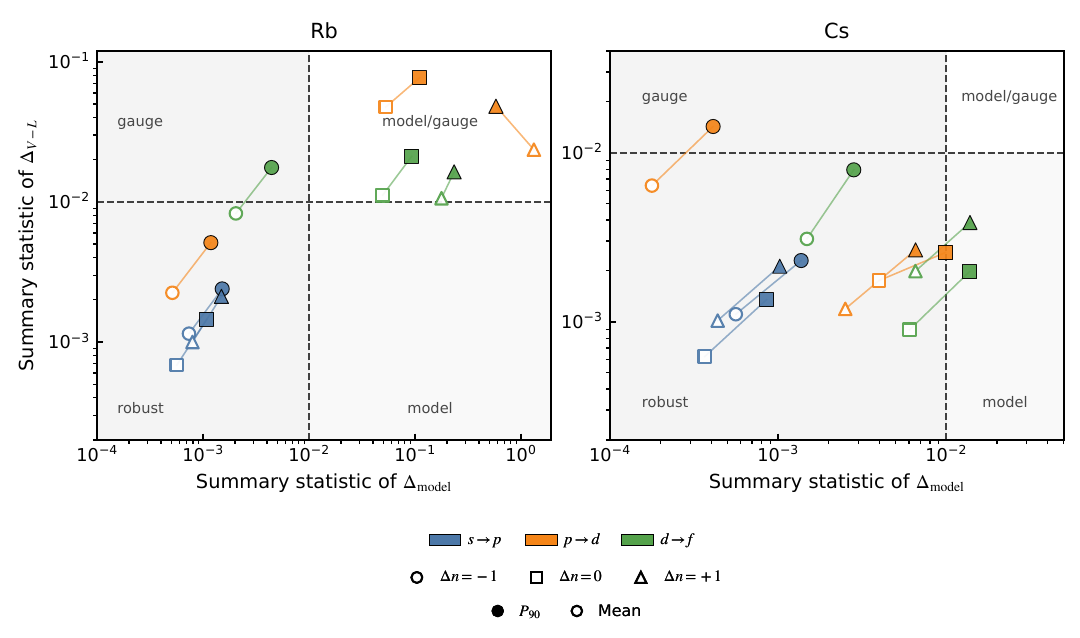}
  \caption{\label{fig:arc_comparison}Summary map combining DFCP--model
differences and internal length-velocity differences. Each point represents
one atom, transition family, and $\Delta n$ group. Filled symbols show the
90th percentile and open symbols show the mean over that group; the thin
line connects the two statistics for the same group. Dashed lines mark the
1\% thresholds used to organize the reliability classification.}
\end{figure*}

Figure~\ref{fig:arc_comparison} places the model-potential difference and the
length-velocity difference on the same plane. The horizontal coordinate
summarizes the difference between the present DFCP matrix elements $D_L$ and the
model-potential values, while the vertical coordinate summarizes the internal
length-velocity difference. Groups in the lower-left quadrant are below the
1\% thresholds in both comparisons. Points to the right of the vertical dashed
line indicate model-sensitive behavior, points above the horizontal dashed
line indicate gauge-sensitive behavior, and points above both thresholds are
sensitive in both senses. The $s$-$p$ groups cluster in the lower-left region 
and are the most stable
for both atoms. The $p$-$d$ and $d$-$f$ groups show larger atom- and
branch-dependent sensitivity, with the most pronounced deviations appearing in
Rb, where several groups enter the gauge-sensitive or model/gauge-sensitive
regions. The complete branch-resolved model-comparison data
underlying the horizontal coordinate are shown in
Fig.~\ref{fig:arc_comparison_full}, and the complete internal-consistency data
underlying the vertical coordinate are shown in
Fig.~\ref{fig:internal_consistency_full}.

The internal-consistency data also show that the length-velocity difference is
the dominant gauge-related diagnostic for the present transitions. The
difference between the CP-corrected length-form matrix element $D_{\rm CP}$
and the ordinary DFCP length-form value $D_L$ remains below 1\% in the full
data set shown in Fig.~\ref{fig:internal_consistency_full}. For example, in
Rb $56d_{5/2}\rightarrow56f_{7/2}$, the length-velocity difference is
1.06\%, whereas the CP-corrected length-form correction is negligible on this
scale. Near the cancellation point of Rb
$57p_{1/2}\rightarrow58d_{3/2}$, the length-velocity difference rises to
63\%, while the CP-corrected length-form correction remains below the percent
level. Because the semiempirical core-polarization potential is not accompanied by a
fully consistent velocity-form effective operator, the exact commutator
relation underlying length-velocity equivalence is not enforced. The
length-velocity difference is therefore used here as an internal
gauge-sensitivity diagnostic rather than as a reference value or a strict
error estimate.

\begin{table*}
\caption{\label{tab:e1_representative}Representative Rydberg-transition $E1$
matrix elements illustrating the reliability classification. $D_L$ is the
ordinary DFCP length-form value, $D_{\rm model}$ is the model-potential value,
$\Delta_{\rm model}=\bigl||D_L|-|D_{\rm model}|\bigr|/|D_L|$, and
$\Delta_{V-L}=\bigl||D_V|-|D_L|\bigr|/|D_L|$. Matrix elements are in a.u.}
\scriptsize
\begin{ruledtabular}
\begin{tabular}{p{0.14\textwidth}p{0.19\textwidth}rrrrp{0.15\textwidth}}
Class & Transition & $|D_L|$ & $|D_{\rm model}|$ & $\Delta_{\rm model}$ & $\Delta_{V-L}$ & Comment \\
\hline
Robust & Rb $40s_{1/2}\to39p_{1/2}$ & 1157.70 & 1157.26 & \num{3.76e-4} & \num{4.37e-4} & Large $n\to n-1$ \\
Robust & Rb $40p_{3/2}\to39d_{5/2}$ & 2984.79 & 2984.45 & \num{1.15e-4} & \num{2.03e-4} & Large $p$-$d$ \\
Cancellation-sensitive & Rb $40p_{3/2}\to40d_{5/2}$ & 18.12 & 18.54 & \num{2.34e-2} & \num{1.94e-2} & Below $D_{\min}$ \\
Model/gauge-sensitive & Rb $56d_{5/2}\to56f_{7/2}$ & 221.40 & 210.03 & \num{5.14e-2} & \num{1.06e-2} & Normal-scale $d$-$f$ \\
Small-denominator cancellation & Rb $57p_{1/2}\to58d_{3/2}$ & 0.0031 & 0.748 & \num{2.40e2} & \num{6.32e-1} & Near cancellation zero \\
Cancellation-sensitive & Rb $30d_{5/2}\to31f_{7/2}$ & 0.321 & 0.0317 & \num{9.01e-1} & \num{6.65e-2} & Small branch \\
Model-sensitive & Cs $88d_{5/2}\to88f_{7/2}$ & 572.61 & 564.01 & \num{1.50e-2} & \num{2.35e-3} & Cs normal-scale \\
\end{tabular}
\end{ruledtabular}
\end{table*}

Table~\ref{tab:e1_representative} lists representative transitions that
illustrate the group-level trends in Fig.~\ref{fig:arc_comparison}. The table
is not intended as a complete database; instead, each example highlights a
typical source of robustness or sensitivity. The Rb
$40s_{1/2}\rightarrow39p_{1/2}$ and
$40p_{3/2}\rightarrow39d_{5/2}$ transitions both have sub-percent agreement
with the model-potential values and small length-velocity differences, and
therefore represent robust large-matrix-element cases.

The largest relative deviations occur primarily in same-$n$ and
$n\rightarrow n+1$ $p$-$d$ and $d$-$f$ branches, but their origins differ. Some
are cancellation spikes. In Rb $57p_{1/2}\rightarrow58d_{3/2}$, the present
length-form value is only $0.0031$ a.u., while the model-potential value is
about $0.748$ a.u. The large relative difference is caused by the small
denominator and indicates that the zero of the radial integral is shifted
between the two calculations. The Rb
$30d_{5/2}\rightarrow31f_{7/2}$ transition is also a small-matrix-element
branch, although it is less singular than the $57p\rightarrow58d$ cancellation
point. The Rb $40p_{3/2}\rightarrow40d_{5/2}$ value lies below the default
$D_{\min}$ scale and is therefore flagged as cancellation-sensitive even though
it is not a near-zero spike.

Other percent-level differences occur for matrix elements of normal size. The
Rb $56d_{5/2}\rightarrow56f_{7/2}$ transition differs from the model-potential
value by 5.1\%. This is not a small-denominator artifact, but reflects
model-dependent changes in the radial wavefunctions. The same transition also
has a 1.06\% length-velocity difference, so it is both model-sensitive and
gauge-sensitive by the criteria above. A more modest normal-scale case is Cs
$88d_{5/2}\rightarrow88f_{7/2}$, for which the model difference is 1.5\%.

These results indicate that the present DFCP calculation provides its most
stable length-form reference values for large, non-cancelling
Rydberg-transition $E1$ matrix elements. For cancellation-sensitive $p$-$d$
and $d$-$f$ transitions, the comparison is better interpreted as a probe of
short-range model dependence than as a single definitive matrix element.

\section{Conclusion}

We have developed and applied a radial DFCP method for Rydberg
states of Rb and Cs. The method treats the closed-shell ionic core using the
DF method, includes exchange with the core explicitly, and incorporates
core-valence correlation through a cutoff core-polarization potential. A radial
Adams-Moulton integration scheme selects the target Rydberg state by its node
number, and a smooth cutoff-radius fit reproduces quantum-defect target
energies with MHz-level residuals up to $n=90$.

The main physical result is that exchange and core polarization can
substantially affect Rydberg-transition $E1$ matrix elements, especially in
$p$-$d$ and $d$-$f$ branches. Radial-integrand and cumulative-integral analyses
show that short-range phase shifts propagate into the outer radial-overlap
region and are amplified when the final matrix element results from
cancellation between large positive and negative contributions. Comparisons
with model-potential matrix elements and internal length-velocity checks show
that many $n\rightarrow n-1$ transitions are robust at the sub-percent level,
whereas same-$n$ and $n\rightarrow n+1$ $p$-$d$ and $d$-$f$ transitions can be
more model-sensitive. The largest relative discrepancies arise either from
near-zero cancellation points or from normal-scale matrix elements with
genuine model dependence of the radial wavefunctions.

The resulting DFCP data provide an independent reference and a practical
reliability assessment for Rydberg-transition $E1$ matrix elements used in
Rydberg-atom field sensing. The comparison does not simply replace
model-potential matrix elements; rather, it identifies which matrix elements
are robust and which should not be regarded as uniquely constrained by the
fitted Rydberg spectrum alone. For field calibration, robust branches are
preferable because matrix-element uncertainty enters directly into the
extracted field amplitude, whereas cancellation-sensitive branches are better
suited as probes of short-range model dependence than as calibration
standards.

\section*{Acknowledgments}

	 We acknowledge support from the Strategic Priority Research Program of the 
   Chinese Academy of Sciences (Grant No. XDB0920403), the National Natural Science Foundation of China (Grant No. 12393821), and the Innovation Fund for Scientific and Technological Personnel of Hainna Province (Grant No. 225RC636).
   
\section*{DATA AVAILABILITY}

	The data that support the findings of this article are not publicly available. The data are available from the authors upon reasonable request.

\appendix

\section{Complete DFCP--model comparison}

\begin{figure*}
  \includegraphics[width=\textwidth]{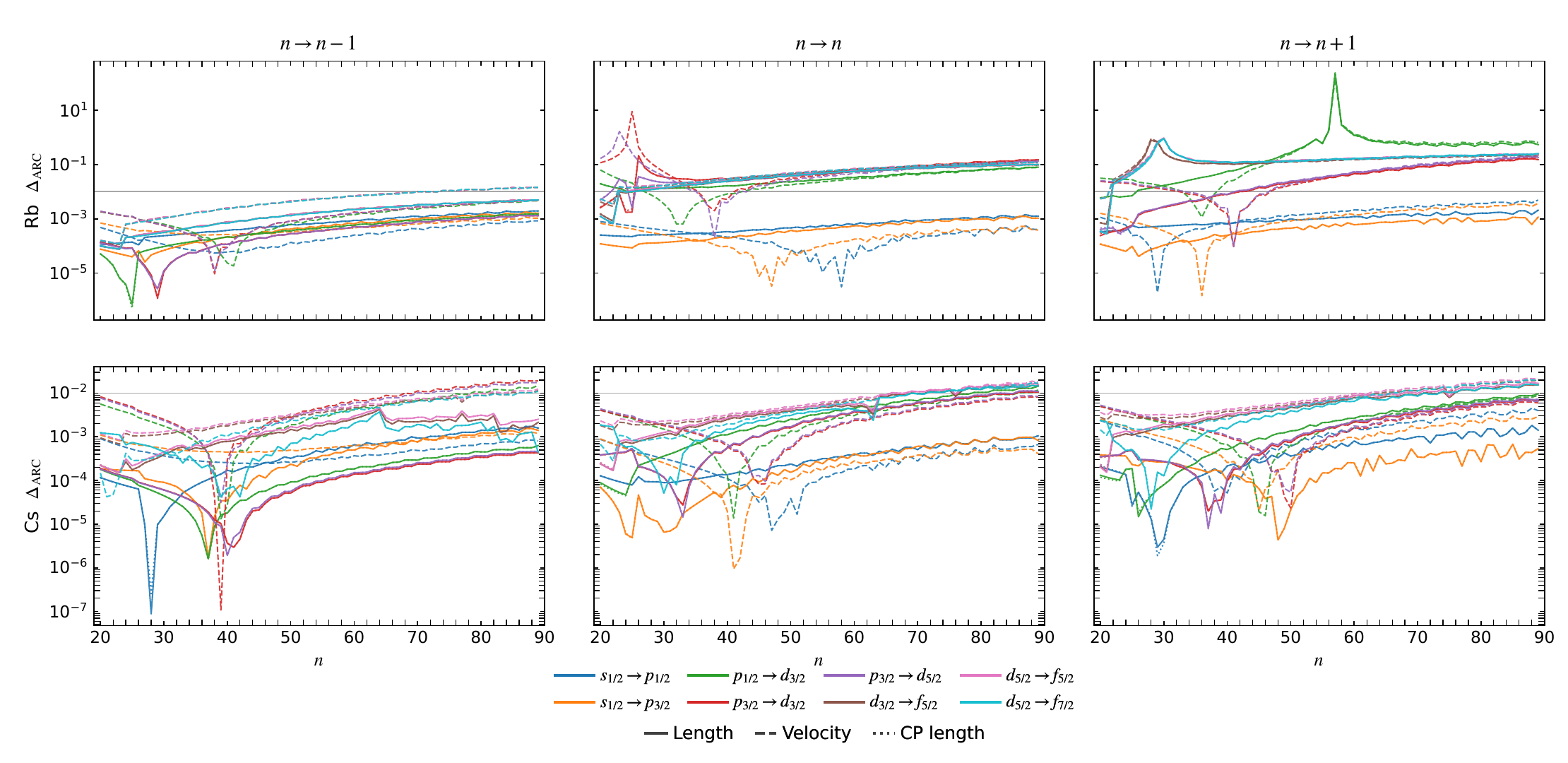}
  \caption{\label{fig:arc_comparison_full}Complete relative differences
  between the present DFCP Rydberg-transition $E1$ matrix elements and
  model-potential values obtained from ARC. The data include
  $\Delta n=-1,0,+1$ branches from $n=20$ to 89. The horizontal line marks
  1\%.}
\end{figure*}

Figure~\ref{fig:arc_comparison_full} gives the complete branch-resolved
DFCP--model comparison underlying the horizontal coordinate of
Fig.~\ref{fig:arc_comparison}. It includes Rb and Cs,
$\Delta n=-1,0,+1$, all branch families, and DFCP matrix elements evaluated
with the length-form, velocity-form, and CP-corrected length-form operators.

\section{Complete internal \texorpdfstring{$E1$}{E1} comparison}

\begin{figure*}
  \includegraphics[width=\textwidth]{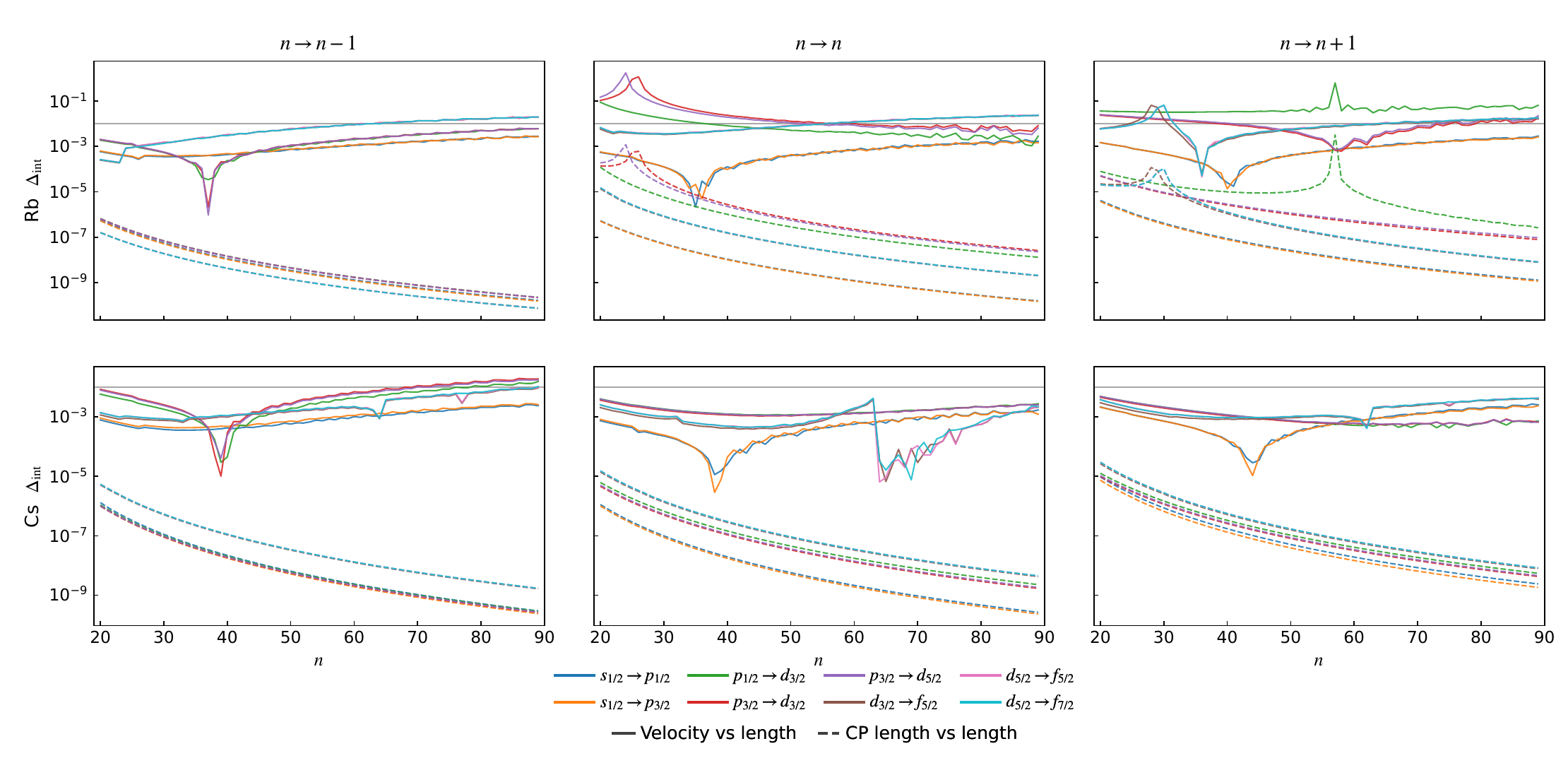}
  \caption{\label{fig:internal_consistency_full}Complete internal DFCP
  comparison between length and velocity forms and between CP-corrected
  length-form and unmodified length-form matrix elements. The unmodified
  length-form value is used as the reference.}
\end{figure*}

Figure~\ref{fig:internal_consistency_full} gives the complete branch-resolved
length-velocity and CP-corrected length-form comparisons underlying the
vertical coordinate and CP-operator discussion associated with
Fig.~\ref{fig:arc_comparison}.

\section{Long-range static RPA limit of the $E1$ operator}
\label{app:cp_length_rpa}

This appendix gives the connection between the CP-corrected length-form
operator in Eq.~\eqref{eq:deff} and the long-range static RPA response of a closed-shell
ionic core.
The derivation uses the static length gauge and retains only the long-range
dipole term of the valence-core Coulomb interaction.

Let the external electric field be $\mathbf E$. With the sign convention used
for the length-form matrix elements in the main text, the direct interaction
of the field with the valence electron is
\begin{equation}
  H_E^{(v)} = \mathbf E\cdot\mathbf r ,
\end{equation}
so that the bare one-electron dipole operator is $\mathbf D=\mathbf r$. The
core $E1$ transition operator is
\begin{equation}
  \mathbf D_c=-\sum_{a\in c}\mathbf r_a ,
\end{equation}
and the corresponding field-core interaction is
\begin{equation}
  H_E^{(c)}=-\mathbf E\cdot\mathbf D_c .
\end{equation}
If $|C\rangle$ is the unperturbed core state and $|N\rangle$ denotes an
excited core state, the first-order core response in the static limit is
\begin{equation}
  |\delta C\rangle =
  \sum_N
  \frac{|N\rangle\langle N|\mathbf E\cdot\mathbf D_c|C\rangle}{E_N-E_C}.
  \label{eq:app_core_response}
\end{equation}
The induced core $E1$ moment is then
\begin{equation}
  \delta\langle \mathbf D_c\rangle =
  \langle \delta C|\mathbf D_c|C\rangle+
  \langle C|\mathbf D_c|\delta C\rangle .
\end{equation}
For a closed-shell core with no preferred orientation, substitution of
Eq.~\eqref{eq:app_core_response} gives the scalar polarizability relation
\begin{equation}
  \delta\langle \mathbf D_c\rangle=\alpha_c \mathbf E ,
  \label{eq:app_core_response_scalar}
\end{equation}
with
\begin{equation}
  \alpha_c=\frac{2}{3}
  \sum_N
  \frac{|\langle C|\mathbf D_c|N\rangle|^2}{E_N-E_C}.
  \label{eq:app_core_polarizability}
\end{equation}

The valence electron at position $\mathbf r$ interacts with the core electrons
through
\begin{equation}
  V_{vc}=\sum_{a\in c}\frac{1}{|\mathbf r-\mathbf r_a|}.
\end{equation}
In the long-range region outside the core, the leading nonspherical term of
the multipole expansion is the dipole term,
\begin{equation}
  V_{vc}^{(1)}
  =
  \sum_{a\in c}\frac{\mathbf r_a\cdot\mathbf r}{r^3}
  =
  -\mathbf D_c\cdot\frac{\mathbf r}{r^3}.
  \label{eq:app_vc_dipole}
\end{equation}
When the external field polarizes the core, Eq.~\eqref{eq:app_vc_dipole}
therefore produces the induced local potential
\begin{equation}
  \delta V_E(\mathbf r)
  =
  -\delta\langle\mathbf D_c\rangle\cdot\frac{\mathbf r}{r^3}
  =
  -\alpha_c\mathbf E\cdot\frac{\mathbf r}{r^3}.
  \label{eq:app_induced_potential}
\end{equation}
The RPA-corrected effective one-electron interaction with the external field
can be written as
\begin{equation}
  H_E^{(v)}+\delta V_E(\mathbf r)
  =
  \mathbf E\cdot\mathbf D_{\mathrm{eff}}(\mathbf r).
\end{equation}
Comparing this equation with Eq.~\eqref{eq:app_induced_potential} gives
\begin{equation}
  \mathbf D_{\mathrm{eff}}(\mathbf r)
  =
  \mathbf r
  -
  \alpha_c\frac{\mathbf r}{r^3}.
  \label{eq:app_deff_long_range}
\end{equation}

The long-range form in Eq.~\eqref{eq:app_deff_long_range} is singular inside
the core and does not describe the short-range nonlocal part of the RPA
response. In DFCP calculations it is therefore regularized with the same
short-range cutoff scale used in the core-polarization potential in
Eq.~\eqref{eq:vcp}. The empirical CP-corrected length-form operator is then
written as
\begin{equation}
  \mathbf D_{\mathrm{eff}}(\mathbf r)
  =
  \mathbf r
  -
  \alpha_c
  \left[1-\exp\left(-\frac{r^6}{\rho^6}\right)\right]^{1/2}
  \frac{\mathbf r}{r^3},
  \label{eq:app_deff_cutoff}
\end{equation}
which is the vector form corresponding to the radial operator in
Eq.~\eqref{eq:deff}.

\bibliography{ref}

@article{PhysRevA.83.052515,
  title = {Measurement of absolute transition frequencies of $^{87}\mathrm{Rb}$ to $\mathit{nS}$ and $\mathit{nD}$ Rydberg states by means of electromagnetically induced transparency},
  author = {Mack, Markus and Karlewski, Florian and Hattermann, Helge and H\"ockh, Simone and Jessen, Florian and Cano, Daniel and Fort\'agh, J\'ozsef},
  journal = {Phys. Rev. A},
  volume = {83},
  issue = {5},
  pages = {052515},
  numpages = {8},
  year = {2011},
  month = {May},
  publisher = {American Physical Society},
  doi = {10.1103/PhysRevA.83.052515}
}

@article{5wyj-9jqj,
  title = {Experimental determination of $S, D, F$, and $G$ energy levels in rubidium},
  author = {True, Timothy M. and Nice, Dillon T. and Mindrup, Jordan C. and Rice, Christopher A. and Perram, Glen P.},
  journal = {Phys. Rev. A},
  volume = {113},
  issue = {3},
  pages = {032811},
  numpages = {9},
  year = {2026},
  month = {Mar},
  publisher = {American Physical Society},
  doi = {10.1103/5wyj-9jqj}
}

@article{PhysRevA.74.054502,
  title = {Rb $nf$ quantum defects from millimeter-wave spectroscopy of cold $^{85}\mathrm{Rb}$ Rydberg atoms},
  author = {Han, Jianing and Jamil, Yasir and Norum, D. V. L. and Tanner, Paul J. and Gallagher, T. F.},
  journal = {Phys. Rev. A},
  volume = {74},
  issue = {5},
  pages = {054502},
  numpages = {4},
  year = {2006},
  month = {Nov},
  publisher = {American Physical Society},
  doi = {10.1103/PhysRevA.74.054502}
}

@article{PhysRevLett.133.233005,
  title = {Ultraprecise Determination of $\mathrm{Cs}(n{S}_{1/2})$ and $\mathrm{Cs}(n{D}_{J})$ Quantum Defects for Sensing and Computing: Evaluation of Core Contributions},
  author = {Shen, Pinrui and Booth, Donald and Liu, Chang and Beattie, Scott and Marceau, Claude and Shaffer, James P. and Pawlak, Mariusz and Sadeghpour, H. R.},
  journal = {Phys. Rev. Lett.},
  volume = {133},
  issue = {23},
  pages = {233005},
  numpages = {7},
  year = {2024},
  month = {Dec},
  publisher = {American Physical Society},
  doi = {10.1103/PhysRevLett.133.233005}
}

@article{dkb4-kb5b,
  title = {Precision measurement of $\mathrm{Cs}(n{F}_{J})$ quantum defects and calculations of scalar and tensor polarizabilities of the $n{S}_{1/2}$, $n{P}_{J}$, $n{D}_{J}$, and $n{F}_{J}$ series},
  author = {Shen, Pinrui and Pawlak, Mariusz and Booth, Donald and Nickerson, Kent and Miladi, Haddad and Sadeghpour, H. R. and Shaffer, James},
  journal = {Phys. Rev. A},
  volume = {112},
  issue = {2},
  pages = {022814},
  numpages = {13},
  year = {2025},
  month = {Aug},
  publisher = {American Physical Society},
  doi = {10.1103/dkb4-kb5b}
}

@article{PhysRevA.93.013424,
  title = {Precision measurement of the ionization energy of Cs i},
  author = {Deiglmayr, Johannes and Herburger, Holger and Sa\ss{}mannshausen, Heiner and Jansen, Paul and Schmutz, Hansj\"urg and Merkt, Fr\'ed\'eric},
  journal = {Phys. Rev. A},
  volume = {93},
  issue = {1},
  pages = {013424},
  numpages = {7},
  year = {2016},
  month = {Jan},
  publisher = {American Physical Society},
  doi = {10.1103/PhysRevA.93.013424}
}

@article{PhysRevA.99.042502,
  title = {Optical spectroscopy of $nP$ Rydberg states of $^{87}\mathrm{Rb}$ atoms with a 297-nm ultraviolet laser},
  author = {Li, Bo and Li, Meng and Jiang, Xiaojun and Qian, Jun and Li, Xiaolin and Liu, Liang and Wang, Yuzhu},
  journal = {Phys. Rev. A},
  volume = {99},
  issue = {4},
  pages = {042502},
  numpages = {7},
  year = {2019},
  month = {Apr},
  publisher = {American Physical Society},
  doi = {10.1103/PhysRevA.99.042502}
}

@article{PhysRevA.100.062515,
  title = {Measurement of the hyperfine coupling constant for $n{S}_{1/2}$ Rydberg states of $^{85}\mathrm{Rb}$},
  author = {Ramos, Andira and Cardman, Ryan and Raithel, Georg},
  journal = {Phys. Rev. A},
  volume = {100},
  issue = {6},
  pages = {062515},
  numpages = {6},
  year = {2019},
  month = {Dec},
  publisher = {American Physical Society},
  doi = {10.1103/PhysRevA.100.062515}
}

@article{PhysRevA.106.052810,
  title = {Hyperfine structure of $n{P}_{1/2}$ Rydberg states in $^{85}\mathrm{Rb}$},
  author = {Cardman, R. and Raithel, G.},
  journal = {Phys. Rev. A},
  volume = {106},
  issue = {5},
  pages = {052810},
  numpages = {5},
  year = {2022},
  month = {Nov},
  publisher = {American Physical Society},
  doi = {10.1103/PhysRevA.106.052810}
}

@article{PhysRevA.110.042815,
  title = {Measurements of the hyperfine structure of $n{P}_{J}$ Rydberg states by microwave spectroscopy in Cs atoms},
  author = {Song, Rong and Bai, Jingxu and Li, Zhenhua and Jiao, Yuechun and Jia, Suotang and Zhao, Jianming},
  journal = {Phys. Rev. A},
  volume = {110},
  issue = {4},
  pages = {042815},
  numpages = {8},
  year = {2024},
  month = {Oct},
  publisher = {American Physical Society},
  doi = {10.1103/PhysRevA.110.042815}
}

@article{cwhz-by5w,
  title = {Applicability of the Dirac-Fock method combined with core polarization in calculations of alkali-metal atoms},
  author = {Bobylev, A. A. and Lopez-Rodriguez, J. J. and Kvasov, P. A. and Reiter, M. A. and Solovyev, D. A. and Zalialiutdinov, T. A.},
  journal = {Phys. Rev. A},
  volume = {113},
  issue = {4},
  pages = {042809},
  numpages = {10},
  year = {2026},
  month = {Apr},
  publisher = {American Physical Society},
  doi = {10.1103/cwhz-by5w}
}

@article{PhysRevA.109.063115,
  title = {$^{229}\mathrm{Th}^{3+}$ as an ionic optical clock for fine-structure-constant variations},
  author = {Yu, Shi-Cheng and Gan, Wen-Ting and Hua, Xia and Tong, Xin and Li, Cheng-Bin},
  journal = {Phys. Rev. A},
  volume = {109},
  issue = {6},
  pages = {063115},
  numpages = {8},
  year = {2024},
  month = {Jun},
  publisher = {American Physical Society},
  doi = {10.1103/PhysRevA.109.063115}
}

@article{stoneTableNuclearMagnetic2005,
  title = {Table of nuclear magnetic dipole and electric quadrupole moments},
  author = {Stone, N.J.},
  year = 2005,
  month = may,
  journal = {At. Data Nucl. Data Tables},
  volume = {90},
  number = {1},
  pages = {75--176},
  issn = {0092640X},
  doi = {10.1016/j.adt.2005.04.001}
}

@article{PhysRevA.94.032503,
  title = {Electric dipole polarizabilities of Rydberg states of alkali-metal atoms},
  author = {Yerokhin, V. A. and Buhmann, S. Y. and Fritzsche, S. and Surzhykov, A.},
  journal = {Phys. Rev. A},
  volume = {94},
  issue = {3},
  pages = {032503},
  numpages = {10},
  year = {2016},
  month = {Sep},
  publisher = {American Physical Society},
  doi = {10.1103/PhysRevA.94.032503}
}

@article{SibalicARCOpensourceLibrary2017,
  title = {ARC: An open-source library for calculating properties of alkali Rydberg atoms},
  shorttitle = {ARC},
  author = {{\v S}ibali{\'c}, N. and Pritchard, J.D. and Adams, C.S. and Weatherill, K.J.},
  year = 2017,
  month = nov,
  journal = {Comput. Phys. Commun.},
  volume = {220},
  pages = {319--331},
  issn = {00104655},
  doi = {10.1016/j.cpc.2017.06.015}
}

@article{robertsonARC30Expanded2021,
  title = {ARC 3.0: An expanded Python toolbox for atomic physics calculations},
  shorttitle = {ARC 3.0},
  author = {Robertson, E.J. and {\v S}ibali{\'c}, N. and Potvliege, R.M. and Jones, M.P.A.},
  year = 2021,
  month = apr,
  journal = {Comput. Phys. Commun.},
  volume = {261},
  pages = {107814},
  issn = {00104655},
  doi = {10.1016/j.cpc.2020.107814}
}

@article{PhysRevA.49.982,
  title = {Dispersion coefficients for alkali-metal dimers},
  author = {Marinescu, M. and Sadeghpour, H. R. and Dalgarno, A.},
  journal = {Phys. Rev. A},
  volume = {49},
  issue = {2},
  pages = {982--988},
  numpages = {0},
  year = {1994},
  month = {Feb},
  publisher = {American Physical Society},
  doi = {10.1103/PhysRevA.49.982}
}

@article{PhysRevA.87.042517,
  title = {Dynamic polarizabilities for the low-lying states of Ca${}^{+}$},
  author = {Tang, Yong-Bo and Qiao, Hao-Xue and Shi, Ting-Yun and Mitroy, J.},
  journal = {Phys. Rev. A},
  volume = {87},
  issue = {4},
  pages = {042517},
  numpages = {11},
  year = {2013},
  month = {Apr},
  publisher = {American Physical Society},
  doi = {10.1103/PhysRevA.87.042517}
}

@article{PhysRevA.69.022509,
  title = {Relativistic many-body calculations of electric-dipole matrix elements, lifetimes, and polarizabilities in rubidium},
  author = {Safronova, M. S. and Williams, Carl J. and Clark, Charles W.},
  journal = {Phys. Rev. A},
  volume = {69},
  issue = {2},
  pages = {022509},
  numpages = {8},
  year = {2004},
  month = {Feb},
  publisher = {American Physical Society},
  doi = {10.1103/PhysRevA.69.022509},
  url = {https://link.aps.org/doi/10.1103/PhysRevA.69.022509}
}

@article{PhysRevA.94.012505,
  title = {Magic wavelengths, matrix elements, polarizabilities, and lifetimes of Cs},
  author = {Safronova, M. S. and Safronova, U. I. and Clark, Charles W.},
  journal = {Phys. Rev. A},
  volume = {94},
  issue = {1},
  pages = {012505},
  numpages = {9},
  year = {2016},
  month = {Jul},
  publisher = {American Physical Society},
  doi = {10.1103/PhysRevA.94.012505},
  url = {https://link.aps.org/doi/10.1103/PhysRevA.94.012505}
}

@article{RevModPhys.82.2313,
  title = {Quantum information with Rydberg atoms},
  author = {Saffman, M. and Walker, T. G. and M\o{}lmer, K.},
  journal = {Rev. Mod. Phys.},
  volume = {82},
  issue = {3},
  pages = {2313--2363},
  numpages = {0},
  year = {2010},
  month = {Aug},
  publisher = {American Physical Society},
  doi = {10.1103/RevModPhys.82.2313},
  url = {https://link.aps.org/doi/10.1103/RevModPhys.82.2313}
}

@ARTICLE{6910267,
  author={Holloway, Christopher L. and Gordon, Joshua A. and Jefferts, Steven and Schwarzkopf, Andrew and Anderson, David A. and Miller, Stephanie A. and Thaicharoen, Nithiwadee and Raithel, Georg},
  journal={IEEE Trans. Antennas Propag.}, 
  title={Broadband Rydberg Atom-Based Electric-Field Probe for SI-Traceable, Self-Calibrated Measurements}, 
  year={2014},
  volume={62},
  number={12},
  pages={6169-6182},
  doi={10.1109/TAP.2014.2360208}}

@article{PhysRevA.101.053432,
  title = {Microwave electrometry via electromagnetically induced absorption in cold Rydberg atoms},
  author = {Liao, Kai-Yu and Tu, Hai-Tao and Yang, Shu-Zhe and Chen, Chang-Jun and Liu, Xiao-Hong and Liang, Jie and Zhang, Xin-Ding and Yan, Hui and Zhu, Shi-Liang},
  journal = {Phys. Rev. A},
  volume = {101},
  issue = {5},
  pages = {053432},
  numpages = {11},
  year = {2020},
  month = {May},
  publisher = {American Physical Society},
  doi = {10.1103/PhysRevA.101.053432},
  url = {https://link.aps.org/doi/10.1103/PhysRevA.101.053432}
}

@article{Gordon,
    author = {Gordon, Joshua A. and Holloway, Christopher L. and Schwarzkopf, Andrew and Anderson, Dave A. and Miller, Stephanie and Thaicharoen, Nithiwadee and Raithel, Georg},
    title = {Millimeter wave detection via Autler-Townes splitting in rubidium Rydberg atoms},
    journal = {Appl. Phys. Lett.},
    volume = {105},
    number = {2},
    pages = {024104},
    year = {2014},
    month = {07},
    issn = {0003-6951},
    doi = {10.1063/1.4890094},
    url = {https://doi.org/10.1063/1.4890094},
}

@article{Holloway2014,
    author = {Holloway, Christopher L. and Gordon, Joshua A. and Schwarzkopf, Andrew and Anderson, David A. and Miller, Stephanie A. and Thaicharoen, Nithiwadee and Raithel, Georg},
    title = {Sub-wavelength imaging and field mapping via electromagnetically induced transparency and Autler-Townes splitting in Rydberg atoms},
    journal = {Appl. Phys. Lett.},
    volume = {104},
    number = {24},
    pages = {244102},
    year = {2014},
    month = {06},
    issn = {0003-6951},
    doi = {10.1063/1.4883635},
    url = {https://doi.org/10.1063/1.4883635},
}

@article{Holloway2017,
    author = {Holloway, Christopher L. and Simons, Matt T. and Gordon, Joshua A. and Dienstfrey, Andrew and Anderson, David A. and Raithel, Georg},
    title = {Electric field metrology for SI traceability: Systematic measurement uncertainties in electromagnetically induced transparency in atomic vapor},
    journal = {J. Appl. Phys.},
    volume = {121},
    number = {23},
    pages = {233106},
    year = {2017},
    month = {06},
    issn = {0021-8979},
    doi = {10.1063/1.4984201},
    url = {https://doi.org/10.1063/1.4984201},
}

@article{PhysRevApplied.21.044025,
  title = {Polarization-insensitive microwave electrometry using Rydberg atoms},
  author = {Cloutman, Matthew and Chilcott, Matthew and Elliott, Alexander and Otto, J. Susanne and Deb, Amita B. and Kj\ae{}rgaard, Niels},
  journal = {Phys. Rev. Appl.},
  volume = {21},
  issue = {4},
  pages = {044025},
  numpages = {6},
  year = {2024},
  month = {Apr},
  publisher = {American Physical Society},
  doi = {10.1103/PhysRevApplied.21.044025},
  url = {https://link.aps.org/doi/10.1103/PhysRevApplied.21.044025}
}

@article{Simons2016,
    author = {Simons, Matt T. and Gordon, Joshua A. and Holloway, Christopher L.},
    title = {Simultaneous use of Cs and Rb Rydberg atoms for dipole moment assessment and RF electric field measurements via electromagnetically induced transparency},
    journal = {J. Appl. Phys.},
    volume = {120},
    number = {12},
    pages = {123103},
    year = {2016},
    month = {09},
    issn = {0021-8979},
    doi = {10.1063/1.4963106},
    url = {https://doi.org/10.1063/1.4963106},
}

@article{PhysRevA.85.062709,
  title = {Nonperturbative $B$-spline $R$-matrix-with-pseudostates calculations for electron-impact ionization of helium},
  author = {Zatsarinny, Oleg and Bartschat, Klaus},
  journal = {Phys. Rev. A},
  volume = {85},
  issue = {6},
  pages = {062709},
  numpages = {10},
  year = {2012},
  month = {Jun},
  publisher = {American Physical Society},
  doi = {10.1103/PhysRevA.85.062709},
  url = {https://link.aps.org/doi/10.1103/PhysRevA.85.062709}
}

@article{Hansen_1993,
doi = {10.1088/0031-8949/1993/T47/001},
url = {https://doi.org/10.1088/0031-8949/1993/T47/001},
year = {1993},
month = {jan},
publisher = {},
volume = {1993},
number = {T47},
pages = {7},
author = {J E Hansen and M Bentley and H W van der Hart and M Landtman and G M S Lister and Y-T Shen and N Vaeck},
title = {The introduction of B-spline basis sets in atomic structure calculations},
journal = {Phys. Scr.}
}

@article{Sapirstein_1996,
doi = {10.1088/0953-4075/29/22/005},
url = {https://doi.org/10.1088/0953-4075/29/22/005},
year = {1996},
month = {nov},
publisher = {},
volume = {29},
number = {22},
pages = {5213},
author = {J Sapirstein and W R Johnson},
title = {The use of basis splines in theoretical atomic physics},
journal = {J. Phys. B: At. Mol. Opt. Phys.}
}

@article{Yu_2025,
doi = {10.1088/1361-6455/ae0a9b},
url = {https://doi.org/10.1088/1361-6455/ae0a9b},
year = {2025},
month = {oct},
publisher = {IOP Publishing},
volume = {58},
number = {19},
pages = {195002},
author = {Yu, Shi-Cheng and Bian, Wu and Li, Cheng-Bin and She, Lei},
title = {Hyperpolarizability effect of Rydberg states in rubidium and cesium atoms},
journal = {J. Phys. B: At. Mol. Opt. Phys.}
}

@article{PhysRevA.109.022810,
  title = {Microwave transitions in atomic sodium: Radiometry and polarimetry using the sodium layer},
  author = {Pawlak, Mariusz and Schoen, Eve L. and Albert, Justin E. and Sadeghpour, H. R.},
  journal = {Phys. Rev. A},
  volume = {109},
  issue = {2},
  pages = {022810},
  numpages = {11},
  year = {2024},
  month = {Feb},
  publisher = {American Physical Society},
  doi = {10.1103/PhysRevA.109.022810},
  url = {https://link.aps.org/doi/10.1103/PhysRevA.109.022810}
}

@article{PhysRevA.110.043114,
  title = {Double, triple, and quadruple magic wavelengths for cesium ground, excited, and Rydberg states},
  author = {Bhowmik, A. and Gaudesius, M. and Biedermann, G. and Blume, D.},
  journal = {Phys. Rev. A},
  volume = {110},
  issue = {4},
  pages = {043114},
  numpages = {19},
  year = {2024},
  month = {Oct},
  publisher = {American Physical Society},
  doi = {10.1103/PhysRevA.110.043114},
  url = {https://link.aps.org/doi/10.1103/PhysRevA.110.043114}
}

@article{PhysRevA.111.053120,
  title = {Forces on alkali Rydberg atoms due to nonlinearly polarized light},
  author = {Bhowmik, A. and Blume, D.},
  journal = {Phys. Rev. A},
  volume = {111},
  issue = {5},
  pages = {053120},
  numpages = {14},
  year = {2025},
  month = {May},
  publisher = {American Physical Society},
  doi = {10.1103/PhysRevA.111.053120},
  url = {https://link.aps.org/doi/10.1103/PhysRevA.111.053120}
}

@book{Johnson2007AtomicStructureTheory,
  author    = {Johnson, Walter R.},
  title     = {Atomic Structure Theory: Lectures on Atomic Physics},
  publisher = {Springer},
  address   = {Berlin, Heidelberg},
  year      = {2007},
  doi       = {10.1007/978-3-540-68013-0}
}

@article{MIGDALEK2020101355,
title = {Model potential study of Rydberg one-electron spectrum of thallium},
journal = {At. Data Nucl. Data Tables},
volume = {135-136},
pages = {101355},
year = {2020},
doi = {https://doi.org/10.1016/j.adt.2020.101355},
author = {J. Migdalek},
}

\end{document}